\documentclass[twocolumn,aps,prl,showpacs,showkeys,assymb,amsmath,superscriptaddress,longbibliography]{revtex4-1}
\usepackage{color}
\usepackage{amsmath}
\usepackage{amssymb}
\usepackage{graphicx}
\usepackage[colorlinks=true,linkcolor=red]{hyperref}
\usepackage[sort&compress]{natbib}
\usepackage[dvipsnames]{xcolor}
\usepackage{soul}

\newcommand{\mh}{\mathcal{H}}
\newcommand{\efb}{E_\text{FB}}

\begin{document}

\title{Compact localized states and flatband generators in one dimension}

\author{Wulayimu Maimaiti}
\affiliation{Center for Theoretical Phyics of Complex Systems, Institute for Basic Science, South Korea.}
\affiliation{IBS School, University of Science and Technology, South Korea.}

\author{Alexei Andreanov}
\author{Hee Chul Park}
\affiliation{Center for Theoretical Phyics of Complex Systems, Institute for Basic Science, South Korea.}

\author{Oleg Gendelman}
\affiliation{Faculty of Mechanical Engineering, Technion, Haifa, Israel}

\author{Sergej Flach}
\affiliation{Center for Theoretical Phyics of Complex Systems, Institute for Basic Science, South Korea.}
\affiliation{New Zealand Institute for Advanced Study, Center for Theoretical Chemistry \& Physics, Massey University, Auckland, New Zealand}

\date{\today}
\begin{abstract}
Flat bands (FB) are strictly dispersionless bands in the Bloch spectrum of a periodic lattice Hamiltonian, recently observed in a variety of photonic and dissipative condensate networks. FB Hamiltonians are finetuned networks, still lacking a comprehensive generating principle. We introduce a FB generator based on local network properties. We classify FB networks through the properties of compact localized states (CLS) which are exact FB eigenstates and occupy $U$ unit cells. We obtain the complete two-parameter FB family of two-band $d=1$ networks with nearest unit cell interaction and $U=2$. We discover a novel high symmetry sawtooth chain with identical hoppings in a transverse dc field, easily accessible in experiments.
Our results pave the way towards a complete description of FBs in networks with more bands and in higher dimensions.
\end{abstract}


\maketitle

Flat band (FB) networks are tight-binding translationally invariant lattices which ensure the existence of one (or several) completely dispersionless bands in the spectrum~\cite{derzhko2015strongly}. FBs have been studied in a number of lattice models in three-dimensional, two-dimensional, and even one-dimensional (1D) settings~\cite{mielke1991ferromagnetism,tasaki1992ferromagnetism,derzhko2006universal,derzhko2010low,hyrkas2013many}, and recently realized experimentally with photonic waveguide networks~\cite{guzman2014experimental,vicencio2015observation,mukherjee2015observation,mukherjee2015observation1,weimann2016transport,xia2016demonstration}, exciton-polariton condensates~\cite{masumoto2012exciton,baboux2016bosonic}, and ultracold atomic condensates~\cite{taie2015coherent,jo2012ultracold}.

At variance with the spatially continuum case of a two-dimensional electron gas with Landau levels of the time-reversal symmetry broken quantum Hall effect, FB networks can co-exist with time reversal symmetry, and essentially rely on destructive interference. The latter is responsible for the existence of compact localized states (CLS). These exact eigenstates to the FB energy  have strictly zero support outside a finite region of the lattice. With one CLS given, the whole CLS set is generated by lattice translations. This set can be orthogonal or non-orthogonal, but still forms a complete basis for the FB Hilbert space. The CLS are classified by the number $U$ of lattice unit cells they occupy~\cite{flach2014detangling}. The presence of a FB signals macroscopic degeneracy and diverging density of states of a corresponding Hamiltonian. Slight perturbations of such a system will in general lift the degeneracy, leading to uniquely defined eigenstates. Emergent transport properties, in turn, are defined by the type of perturbation. The zero width of the FB calls for non-perturbative effects of the weakest perturbations like disorder or many-body interactions. Thus FB models are high-symmetry cases in a general control parameter space of perturbed lattice Hamiltonians, at which qualitatively different physical phases of matter meet, similar to quantum phase transition points~\cite{sachdev2007quantum}. Examples of such nontrivial and abrupt changes of the wavefunction properties of perturbed FB systems are the appearance of flatband ferromagnetism for many-body interacting fermions~\cite{derzhko2015strongly,mielke1991ferromagnetism,tasaki1992ferromagnetism}, energy dependent scaling of disorder-induced localization length~\cite{flach2014detangling,leykam2016localisation}, singular mobility edges with quasiperiodic potentials~\cite{bodyfelt2014flatbands,danieli2015flatband}, and Landau-Zener Bloch oscillations in the presence of external fields~\cite{khomeriki2016landau}.

It is highly desirable to introduce a clear classification of FBs, and to have local FB testing routines which tell FBs and non-FBs apart avoiding potentially costly band structure calculations. But the most important missing item is a systematic FB generator which allows to obtain all FB models within a given class. Several approaches to construct FB networks have been proposed using graph theory~\cite{mielke1991ferromagnetism}, local cell construction~\cite{tasaki1992ferromagnetism}, so-called "\emph{Origami rules}" in decorated lattices~\cite{dias2015origami}, and repetitions of mini-arrays~\cite{morales2016simple}. None of them starts with a classification of FBs, and can be therefore considered at best as a partial accomplishing of a generator which lacks completeness. In addition, a number of FB models have been identified by intuition or simply accidentally~\cite{vidal1998aharonov}.

A first attempt to classify FBs through the properties of CLS was published in Ref.~\cite{flach2014detangling}. The observation was that for $U=1$ the CLS set forms an orthogonal complete FB basis, with the possibility to detangle the CLS from the rest of the lattice. The inverse procedure - taking {\sl any} lattice, assigning a set of $\nu$ detangled CLS states with energies $\epsilon_\nu$ to each unit cell of the lattice, and finally performing the inverse entangling procedure of mixing the CLS states with the states from each unit cell - leads to the most general $U=1$ FB generator for arbitrary lattice dimension, arbitrary number of bands, and arbitrary number of FBs amongst them~\cite{flach2014detangling}. However, for all $U > 1$ cases - for various reasons the more interesting and nontrivial ones - the inverse detangling method fails, since CLS states are not anymore orthogonal. Therefore, we are in need of a different approach. In this work we present the first nontrivial FB generator for $U=2$ in one dimension. We also define a simple local FB tester routine.

We consider a one-dimensional ($d=1$) translationally invariant lattice with $\nu>1$ lattice sites per unit cell $n$ and the wave function $\Psi=(...,\vec{\psi}_{n-1},\vec{\psi}_n,...)$, where the individual vectors $\vec{\psi}_n$ have elements  $\psi_{na}$, and  $a=1,...,\nu$ labels the sites inside the unit cell. The time-independent Schr\"odinger equation on such a network is given by
\begin{equation}
	\label{fncls:seqn}
    \mh\Psi = E\Psi \;,
\end{equation}
where $\mh$ is the Hamiltonian matrix of the network, and $E$ is the eigenenergy. Discrete translational invariance assumes that $\mh$ is invariant under shifts $n\rightarrow n+p$. With the help of the Floquet-Bloch theorem the eigenstates of~\eqref{fncls:seqn} can be represented as $\vec{\psi}_{n}=\sum_k\mathrm{e}^{ikn}\vec{u}(k)$ where $\vec{u}_\mu(k)$ is the polarization vector of the $\mu$th band with $\mu=1,2,\dots,\nu$ and $k$ is a wave number. The eigenvalues form $\nu$ bands $E_\mu(k) = E_\mu(k+2\pi)$. 
For the purpose of clear classification we decompose $\mh$ into a block matrix:
\begin{equation}
	\label{fncls:hopping}
    \sum_{m=-\infty}^\infty H_m\vec{\psi}_{n+m}=E\vec{\psi}_n \; ,
\end{equation}
where the $\nu\times\nu$ matrices $H_m=H_{-m}^\dagger$ describe the hopping (tunneling)  between sites from unit cells at distance $m$~\footnote{Note that $H_0$ is Hermitian, while $H_m$ with $m\neq0$ are not in general.}. We further classify networks according to the largest hopping range $m_c$: $H_m\equiv0$ for $|m|>m_c\geq1$. Note that $H_0$ describes intracell connectivities and $H_{m \neq 0}$ intercell links.

\begin{figure}
    \includegraphics[clip,width=\columnwidth]{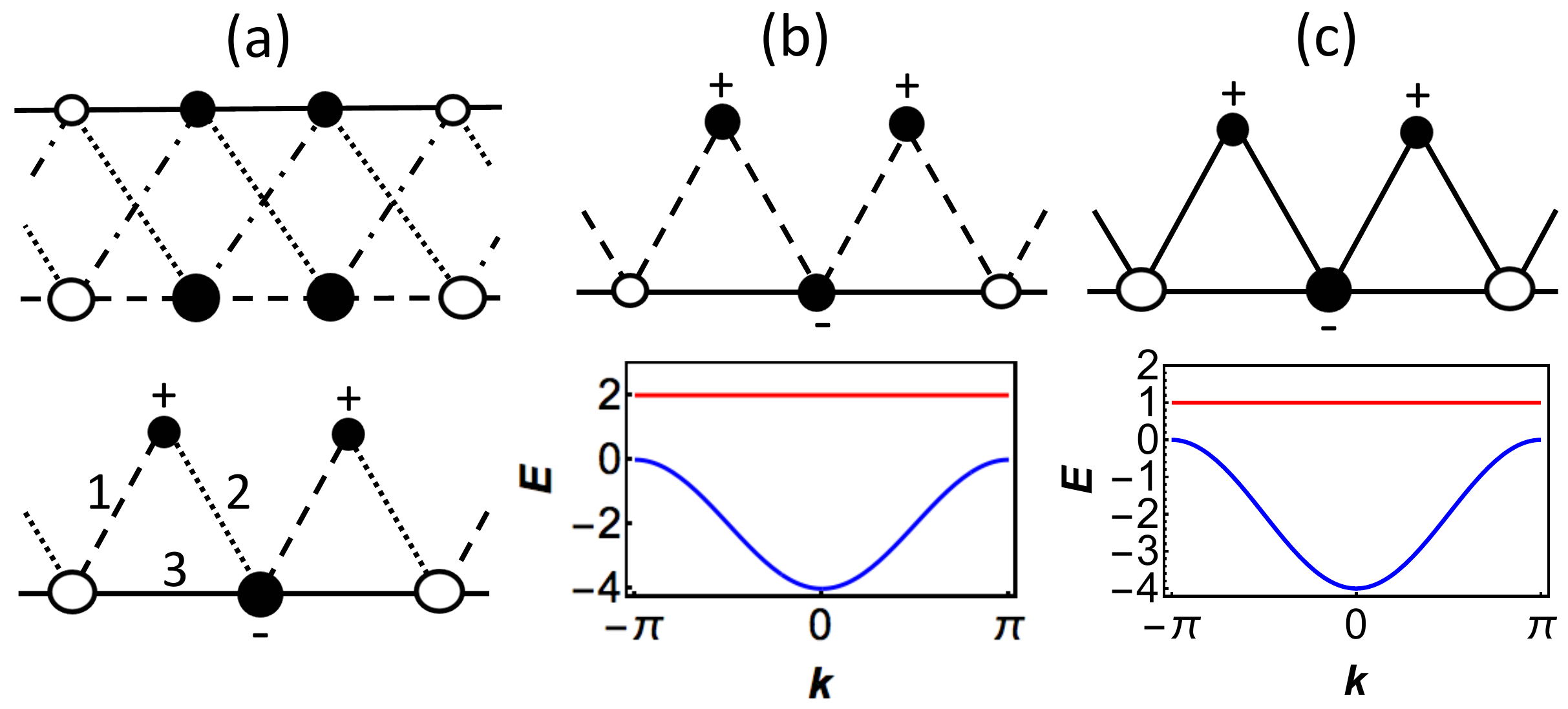}
    \protect\caption{(color online) 
	(a) Top: canonical $\nu=2$ chain for $U=2$. Circles - lattices sites (different sizes correspond to different onsite energies). Lines - hopping connections (different lines correspond to different hopping strengths). Filled circles - location of a CLS. Bottom: generalized sawtooth chain after basis rotation (see text for details). Signs indicate signs of the CLS amplitudes. (b) The known sawtooth ST1 chain. (c) New sawtooth ST2 chain. Top of (b,c): lattice structure, Bottom of (b,c): band structure.
}
    \label{fig1}
\end{figure}

A \emph{compact localized state} (CLS) is a solution of~\eqref{fncls:hopping} with $\vec{\psi}_n\neq0$ only on the smallest possible finite number $U$ of adjacent unit cells and zero everywhere else~\cite{flach2014detangling}. The corresponding eigenenergy is denoted as $\efb$. If such an eigenstate exists, then its translations along the lattice are also eigenstates, leading to a macroscopic degeneracy of $\efb$. The resulting band is flat, i.e. $E_\mu(k)=\efb$ is independent of $k$. We arrive at three essential control parameters which classify FB networks: the hopping range $m_c$, the number of bands $\nu$ and the CLS class $U$. The lattice and band structure of the cross-stitch lattice with $U=1$, $\nu=2$, $m_c=1$ was reported in ~\cite{flach2014detangling} and corresponds to $H_0=\left( \begin{array}{cc} 0 & 0 \\ 0 & 0 \end{array} \right) $ and  $H_1=-\left( \begin{array}{cc} 1 & 1 \\ 1 & 1 \end{array} \right)$. In Fig.~\ref{fig1}(b) the sawtooth lattice (ST1) with $U=2$, $\nu=2$ and $m_c=1$ is shown~\cite{flach2014detangling} together with its CLS and band structure, with $H_0= -\left( \begin{array}{cc} 0 & \sqrt{2} \\ \sqrt{2} & 0 \end{array} \right) $ and  $H_1= -\left( \begin{array}{cc} 0 & \sqrt{2} \\ 0 & 1 \end{array} \right) $. ST1 and its FB were recently experimentally probed with photonic waveguide lattices~\cite{weimann2016transport}.

\begin{figure}
    \includegraphics[clip,width=\columnwidth]{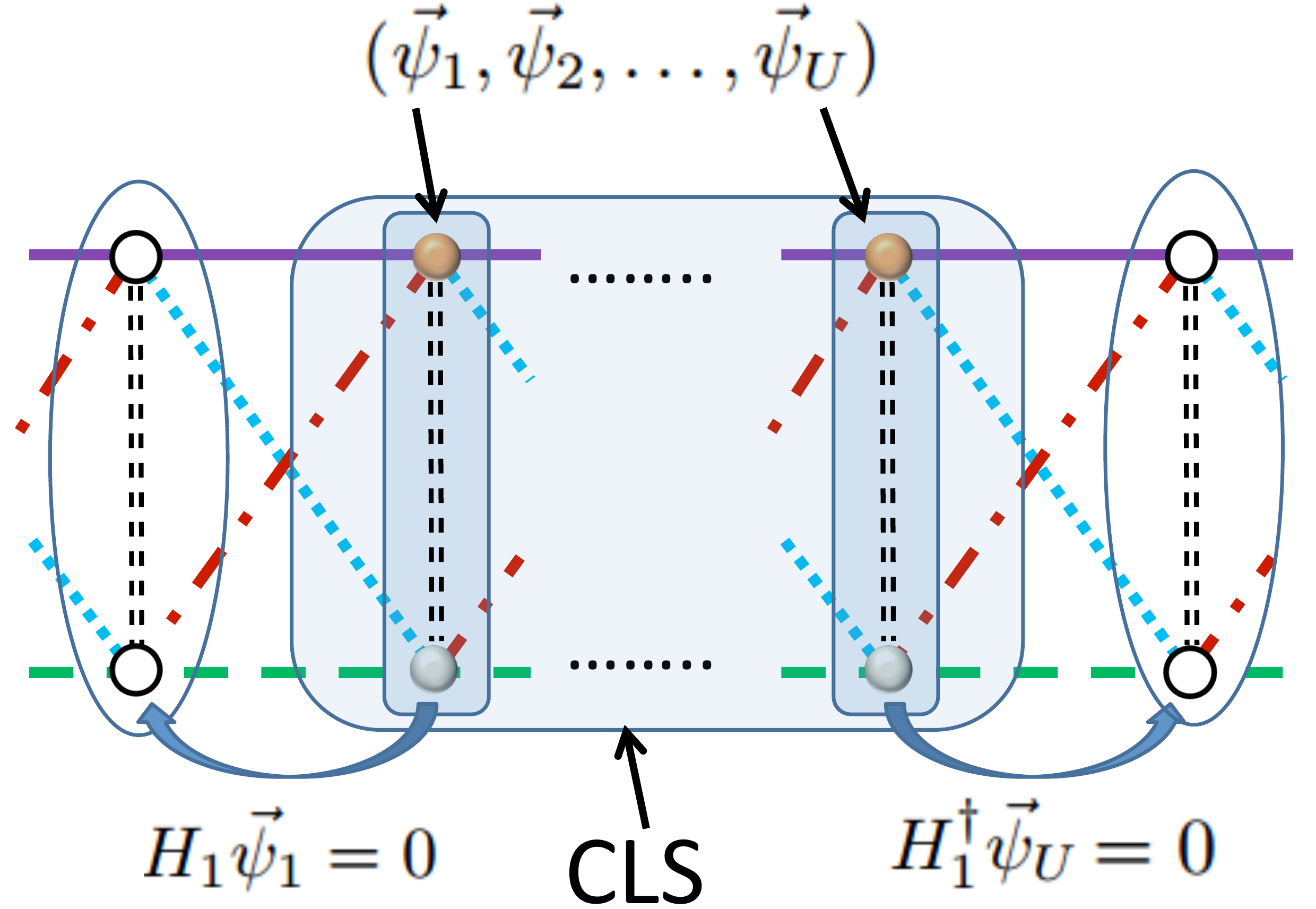}
    \protect\caption{(color online) Schematics of the compact localized state with $m_c=1$ and $\nu=2$.}
    \label{fig2}
\end{figure}

The existence of CLS in a FB lattice can now be used to design a simple local test routine as to whether a given network has a FB of class $U$ or not. Consider the $U\times U$ block matrix
\begin{gather}
    \mh_U=\left(\begin{array}{cccccc}
        H_0 & H_1 & H_2 & H_3 & \dots & H_U\\
        H_1^\dagger & H_0 & H_1 & H_2 & \dots & H_{U-1}\\
        \vdots & \ddots & \ddots & \ddots & \ddots & \vdots\\
        \vdots & ~ & ~ & ~ & ~ & \vdots\\
        H_{U-1}^\dagger & \dots & H_2^\dagger & H_1^\dagger & H_0 & H_1\\
        H_U^\dagger & \dots & H_3^\dagger & H_2^\dagger & H_1^\dagger & H_0
    \end{array}\right)
    \label{tcls:block}
\end{gather}
and an eigenvector $(\vec{\psi}_1,\vec{\psi}_2,\dots,\vec{\psi}_U)$ with eigenvalue $\efb$ such that
\begin{equation}
    \sum_{m=-m_c}^{m_c} H_m \vec{\psi}_{p+m} = 0 \;,\; \vec{\psi}_{l \leq 0} = \vec{\psi}_{l > U} = 0
    \label{test}
\end{equation}
for all integers $p$ with $-m_c+1 \leq p \leq 0$ and $U+1 \leq p \leq U+m_c$. Similar equations hold for $H_m^\dagger$. These two sets of equations ensure $\vec{\psi}_{l \leq 0} = \vec{\psi}_{l > U} = 0$. Then the Hamiltonian is FB to class $U$. As an example, consider $m_c=1$, see Fig.~\ref{fig2}. The corresponding condition simplifies to $H_1^\dagger \vec{\psi}_U = H_1 \vec{\psi}_1 = 0$. Given a network Hamiltonian, and successively increasing the test value for $U=1,2,...$ we arrive at a systematic procedure to identify a FB model with any finite class $U$. Given one computed CLS, the whole set of CLS vectors is generated by discrete translations along the lattice. This set is linearly independent if the set of vectors $\{ \vec{\psi}_1,..,\vec{\psi}_U\}$ is linearly independent~\cite{supp}. It follows that the CLS set for $U=1,2$ is always linearly independent~\cite{supp}. A linearly independent CLS set has dimension equal to the dimension of the Bloch eigenvectors of the FB, and therefore completely spans the FB Hilbert subspace. The Bloch polarization vectors $\vec{u} (k)$ of the FB can then be obtained by computing $\vec{u} (k) \sim \sum_{l=1}^U \vec{\psi}_l e^{i(U-1)k}$~\cite{supp}.

With that we arrive at our core result - a novel systematic local FB generator based on CLS properties. Without loss of generality we will use a \emph{canonical form} of $\mh$: A unitary transformation on each unit cell will diagonalize $H_0$ sorting its diagonal elements (eigenvalues) $H_{\mu\mu}$ monotonically increasing with $\mu$. A trivial gauge $\mh \rightarrow \mh+\zeta \mathcal{I}$ (with $\mathcal{I}$ the identity matrix) sets $H_{11} = 0$, and a subsequent rescaling $\mh \rightarrow \kappa \mh$ ensures $H_{22} = 1$  (the case of a completely degenerate $H_0$ will be treated separately). For convenience we set $m_c=1$ which corresponds to nearest neighbour hopping and is one of the most typical cases considered both experimentally and theoretically. Then we have to find those $\nu\times\nu$ matrices $H_0,H_1$ which solve the following set of equations for $1 \leq l \leq U$:
\begin{eqnarray}
	\label{mc1-1}
    H_1^\dagger\vec{\psi}_{l-1} + H_0\vec{\psi}_l + H_1\vec{\psi}_{l+1}  =  \efb\vec{\psi}_l\;,\\
    \label{mc1-2}
    H_1^\dagger\vec{\psi}_1 = H_1\vec{\psi}_U=0\;,\; \vec{\psi}_0 = \vec{\psi}_{U+1} = 0\;.
\end{eqnarray}
Choosing a set of $H_0,H_1$ we solve the eigenvalue problem~\eqref{mc1-1}) under the constraint of~\eqref{mc1-2} which makes $H_1$ singular and $\vec{\psi}_1$ and $\vec{\psi}_U$ the left and right eigenvectors of the zero mode(s) of $H_1$.

Let us choose the simplest yet nontrivial case of two bands $\nu=2$ which completely fixes the non-degenerate matrix $H_0$:
\begin{equation}
	\label{nu=2}
    H_0=\left(\begin{array}{cc}
        0 & 0\\
        0 & 1
    \end{array}\right)
    \;,\;
    H_1=\left(\begin{array}{cc}
        a & b\\
        c & d
    \end{array}\right).
\end{equation}
Since $H_1$ is singular and of size $2$, it has exactly one zero mode, and  can be parametrized with $H_1 =\alpha\vert\theta,\delta\rangle\langle\varphi,\gamma\vert$ as follows:
\begin{equation}
	\label{res:H1-angles}
    \begin{aligned}H_1 = \alpha\left(
        \begin{array}{cc}
            \cos\theta\cos\varphi & e^{i\gamma}\cos\theta\sin\varphi\\
            e^{-i\delta}\sin\theta\cos\varphi & e^{-i\left(\delta-\gamma\right)}\sin\theta\sin\varphi
        \end{array}\right) \;,
    \end{aligned}
\end{equation}
The prefactor $\alpha = |\alpha | e^{i\phi_\alpha}$ can be complex, and $\vert\varphi,\gamma\rangle$ and $\vert\theta,\delta\rangle$ are the left and right eigenvectors of the non-zero eigenvalue of $H_1$~\cite{supp}. The upper plot in Fig.~\ref{fig1}(a) illustrates this canonical network structure. A rotation of the unit cell basis by an angle $\omega$ shifts the angles $\theta \rightarrow \theta +\omega$ and $\varphi \rightarrow \varphi+\omega$, and modifies $H_0=\left( \begin{array}{cc} \cos^2 \omega & \cos \omega \sin \omega \\ \cos \omega \sin \omega & \sin^2 \omega \end{array} \right)$. Therefore the canonical structure can be always mapped onto a \textsl{generalized sawtooth} (gST) chain with three different hoppings $t_{1,2,3}$ per triangle, and an onsite energy detuning (see bottom of Fig.\ref{fig1}(a) and~\cite{supp} for details).

We test our generator with the known solutions for $U=1$ (Fig.~\ref{fig1}(a) in Ref.~\cite{flach2014detangling}). Equations~(\ref{mc1-1}-\ref{mc1-2}) reduce to $H_0 \vec{\psi}_1= \efb \vec{\psi}_1$ and $H_1\vec{\psi}_1=H_1^{\dagger}\vec{\psi}_1=0$. Then the FB energy is $\efb=0$ or $\efb=1$. For $\efb=0$ it follows $\theta=\pi/2$ or $3\pi/2$ and $\varphi=\pi/2$ or $3\pi/2$. Respectively for $\efb=1$ we find $\theta=0$ or $\pi$ and $\varphi=0$ or $\pi$. The canonical form of $H_1$ has exactly one nonzero element on the diagonal, e.g.  for $\efb=0$ it  is $H_1=\left(\begin{array}{cc} 0 & 0\\ 0 & |\alpha|e^{i\phi_{\alpha}} \end{array}\right)$. We therefore obtain the detangled structure of the cross-stitch lattice (cf. Fig.2(a) in Ref.~\cite{flach2014detangling}). The dispersive band energy is given by $E(k)=C+2|\alpha|\cos\left(k+\phi_{\alpha}\right)$ where $C=0$ for $\efb=1$ and $C=1$ for $\efb=0$. The case of degenerate $H_0\equiv 0$ does not change the structure of $H_1$ and leads to $\efb=0$ and $C=0$. Interestingly the cross-stitch lattice family in Ref.~\cite{flach2014detangling} was characterized by three parameters - the location of the flat band energy, the width of the dispersive band, and an overall gauge. Here we obtain a four-dimensional control parameter space. The first three - the overall gauge $\zeta$, the rescaling $\kappa$ and the band width control $|\alpha|$ reproduce the findings from Ref.~\cite{flach2014detangling}. The additional fourth control parameter is the phase $\phi_\alpha$. It corresponds to a time-reversal symmetry breaking effective magnetic field in one dimension, and completes the class of $m_c=1,\nu=2,U=1$ FB lattices. Remarkably there is another hidden $U=1$ case with two flat bands for which $H_1$ has precisely one nonzero element on one of the two off-diagonals: $\theta=0, \varphi=\pi/2$ or $\theta=\pi/2, \varphi=0$. To observe that one has also to redefine the unit cell \cite{supp}.

We proceed to the nontrivial $U=2$ case. In this case the Hamiltonian $\mh_{U=2}$ is a $2\times2$ block matrix
\begin{equation}
	\label{u2:mh_u2}
    \mh_2 = \left(\begin{array}{cc}
        H_0 & H_1\\
        H_1^\dagger & H_0
    \end{array}\right)
\end{equation}
where $H_0$ is given by~\eqref{nu=2} and $H_1$ is given by~\eqref{res:H1-angles}.

\begin{figure}[!tbp]
    \centering
    \includegraphics[clip,width=1\columnwidth]{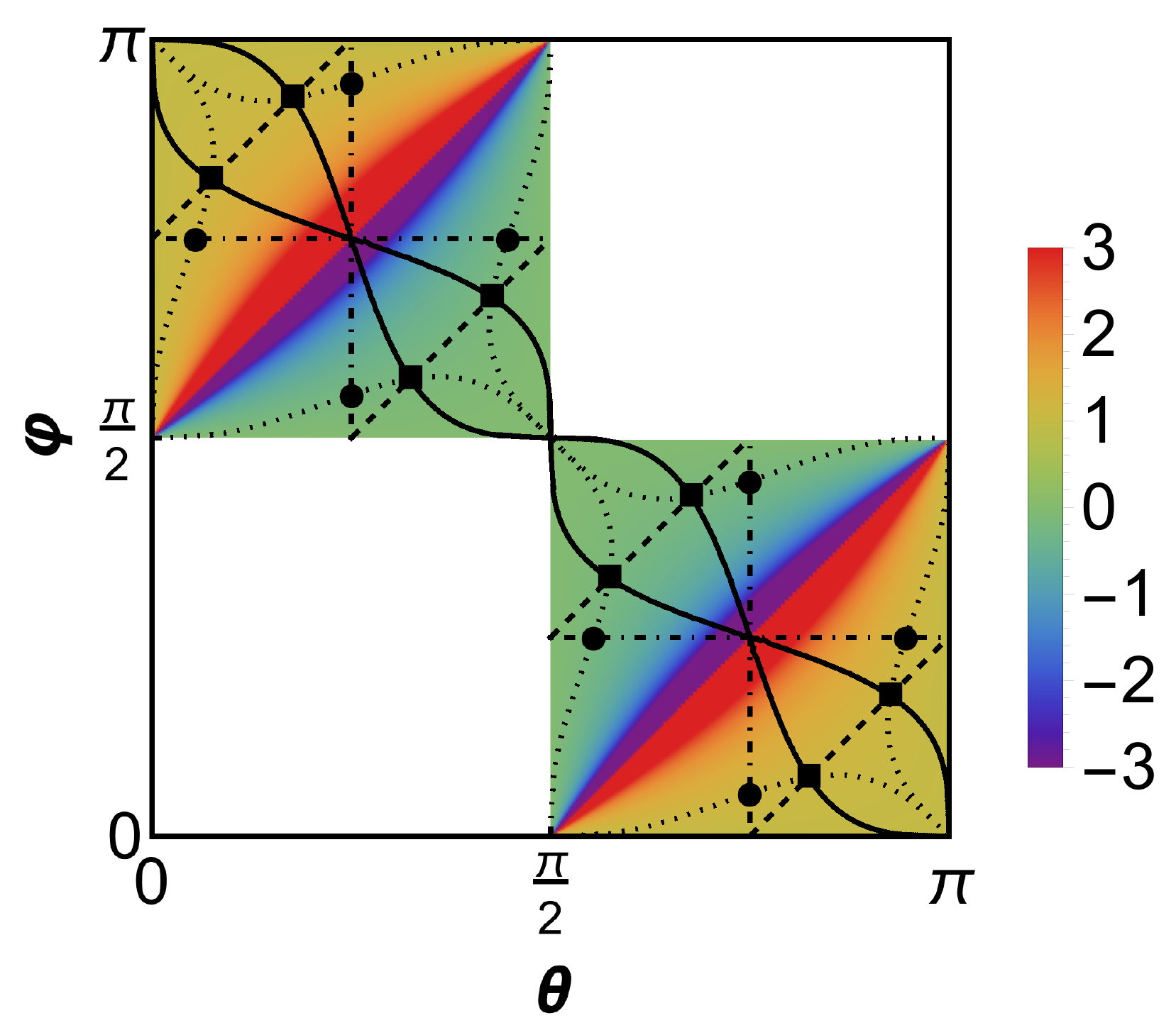}
    \protect\caption{(Color online) 
The flatband energy $\efb \left(\theta,\varphi\right)$ for $m_c=1,\nu=2,U=2$. The colored squares host FB networks, while the white ones do not. The color code shows the energy of the FB. Dashed-dotted lines - same onsite energies in gST chain. Dotted lines - $t_1=t_2$ in gST chain. Dashed lines - $t_2=t_3$ in gST chain. Solid lines - $t_1=t_3$ in gST chain. Filled circles: ST1 chain in Fig.~\ref{fig1}(b). Filled squares : ST2 chain in Fig.\ref{fig1}(c).
}
    \label{fig3}
\end{figure}

The equations~(\ref{mc1-1}-\ref{mc1-2}) read
\begin{align}
	\label{eq:eigenvalue_eq2}
    H_0\vec{\psi}_1 + H_1 \vec{\psi}_2 & = & \efb\vec{\psi}_1 & \;, &
    H_1\vec{\psi}_1 & = & 0 \;,\\
    \label{eq:cls_con2}
    H_1^\dagger\vec{\psi}_1 + H_0\vec{\psi}_2 & = & \efb\vec{\psi}_2 & \;, &
    H_1^\dagger\vec{\psi}_2 & = & 0 \;.
\end{align}
The details of solving the above equations are given in~\cite{supp}. The final result reads:
\begin{equation}
    \delta  =  \gamma,\,\,
    |\alpha|  =  \frac{\sqrt{-\sin(2\theta)\sin(2\varphi)}}{|\sin(2(\theta-\varphi))|}.
    \label{u2:sol}
\end{equation}
The solutions~\eqref{u2:sol} and the Hamiltonian $\mh$ are invariant under the transformation $\{\varphi \rightarrow \varphi+p \pi\;,\; \theta \rightarrow \theta + q \pi\;, \phi_{\alpha} \rightarrow \phi_\alpha+(p+q)\pi\}$ with $p,q$ being integers. The irreducible angle parameter space therefore reduces to $0 \leq \varphi,\theta \leq \pi$. Since $|\alpha|$ is real, the solutions only exist for $0\le\theta\le\frac{\pi}{2}\cap\frac{\pi}{2}\le\varphi\le\pi$ or $\frac{\pi}{2}\le\theta\le\pi\cap0\le\varphi\le\frac{\pi}{2}$, i.e. two disjoint regions shown for the flatband energy $\efb$ in Fig.~\ref{fig3}. The corresponding band structure is given by~\cite{supp}
\begin{align}
	\label{efbu2}
    \efb & =  \frac{\cos(\theta)\cos(\varphi)}{\cos(\theta-\varphi)},\\
    \label{e(k)u2}
    E(k) & =  \frac{\sin(\theta) \sin(\varphi)}{\cos(\theta-\varphi)} + 2 | \alpha | \cos(\theta-\varphi)\cos(k+\phi_\alpha).
\end{align}

The bandwidth $\Delta_w$ of the dispersive band  is given by
\begin{equation}
	\label{bw}
    \Delta_w = 2 \frac{\sqrt{-\sin(2\theta)\sin(2\varphi)}}{| \sin(\theta - \varphi) |}
\end{equation}
and is always bounded $|\Delta_w| \leq 2$.

\begin{figure}[!tbp]
    \centering
    \includegraphics[clip,width=1\columnwidth]{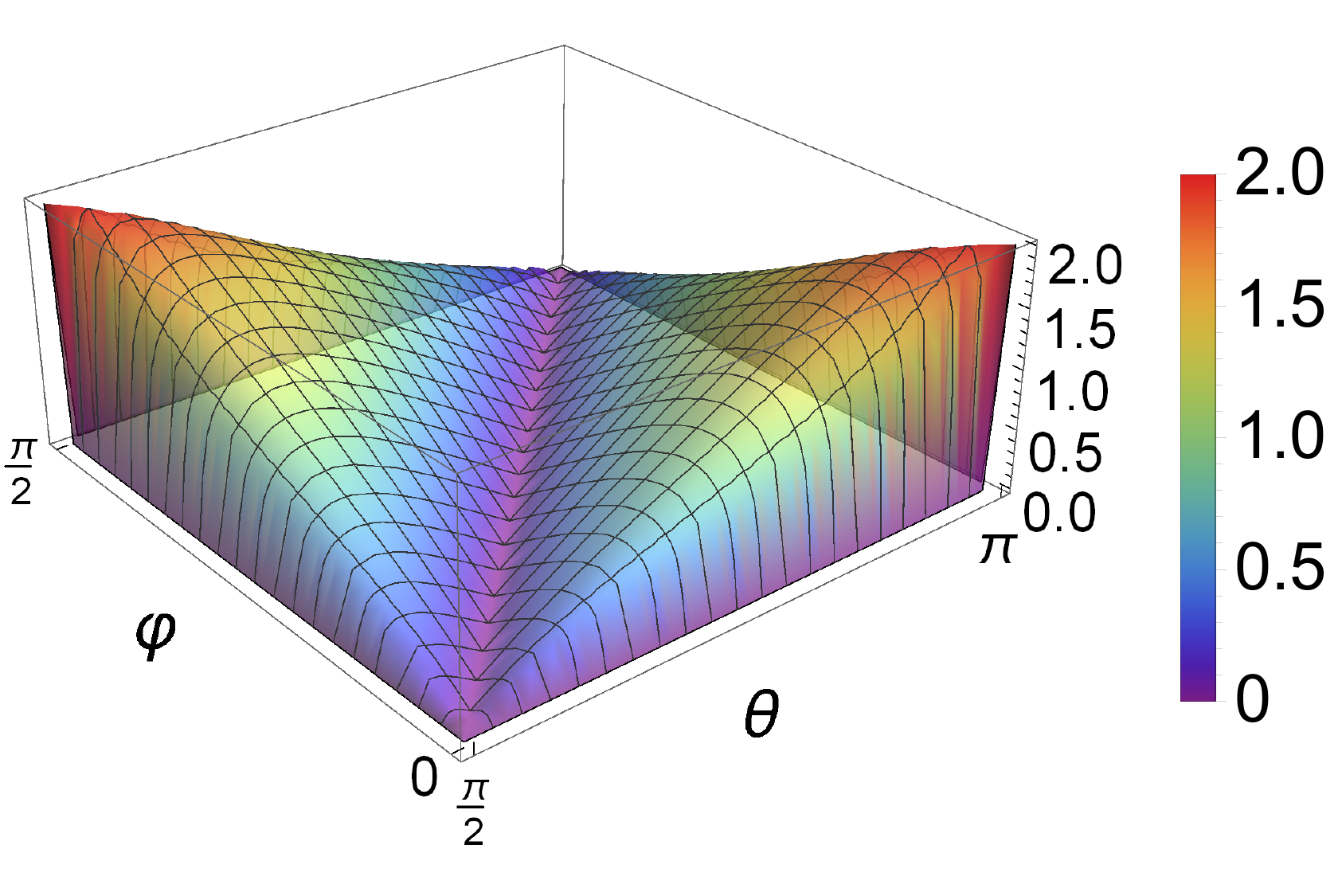}
    \protect\caption{The ratio of the dispersive band width to the distance between the flat band and the dispersive band center $\Delta_w/\Delta E$ versus  $\left(\theta,\varphi\right)$ for $m_c=1,\nu=2,U=2$ in one irreducible quadrant. The color code shows the value of the ratio.}
    \label{fig4}
\end{figure}

The flatband energy is always gapped away from the dispersive band by a gap $\Delta_g= \Delta E - \frac{\Delta_w}{2} $ with $\Delta E$ being the distance between the flat band energy and the dispersive band center (except for few isolated points discussed below). The ratio $\Delta_w/ \Delta E$ is shown in Fig.~\ref{fig4}. This ratio is zero for $\theta = \pi/2+\varphi$ . There the FB energy is gapped infinitely far away from the dispersive band. Using a proper rescaling parameter $\kappa$ and a gauge $\zeta$ we can always renormalize the band gap to a finite number, at the expense of flattening the dispersive band. This special line corresponds to the case of degenerate $H_0$ and two flat bands of class $U=1$  (see \cite{supp} for details). On the boundary lines $\varphi,\theta = 0,\pi/2,\pi$ the band width $\Delta_w$ strictly vanishes, reducing the problem to a trivial $H_1=0$ case with two flat bands of class $U=1$. One exception are the points $\{ \theta = \pi-\varphi \;,\; \varphi=0,\pi/2,\pi \}$ where the band width $\Delta_w$ stays finite but the gap $\Delta_g$ vanishes. Here the flatband becomes of class $U=1$ and touches the dispersive band of finite width (see Fig.~\ref{fig4}).

The class $U=2$ is the largest possible irreducible CLS for $\nu=2,\ m_c=1$ networks, as can be straightforwardly checked using the above generator construction~\cite{supp}~\footnote{This conclusion can also be verified e.g. using a band structure calculation.}. 
Moreover, we find all one-parameter families of gST chains for which either the onsite energies are equal (dashed-dotted lines in Fig.~\ref{fig3}), or a pair of the hoppings $t_{1,2,3}$ is equal (solid, dashed and dotted lines in Fig.~\ref{fig3}). The known ST1 chain is given by the intersection of dotted and dashed-dotted lines - two hoppings are equal $t_1=t_2 \neq t_3$, and the onsite energies are equal. We discover a novel intersection point (square symbols in Fig.~\ref{fig3}) where all three hoppings are equal $t_1=t_2=t_3$  (but the onsite energies differ). This is a new ST2 chain (Fig.~\ref{fig1}(c)), which should be easily realized experimentally - simple geometry allows to make all hoppings equal, an external dc bias will finetune the onsite energy differences, and the CLS is easily addressed having identical absolute amplitude values on occupied sites.

To summarize, we introduced a novel flatband generator - a systematic approach to construct FB networks and  a complete classification of flatband lattices using compact localized (eigen)states of class $U$. This method 
allows to construct flat band networks with predefined properties: the number of flat bands, their position with respect to the regular band, etc. We construct the whole FB family of two-band networks with nearest neighbor hopping. 
We propose a simple testing routine which allows to conclude whether a given lattice is a FB network or not, without computing the band structure. The boundaries of the existence of the $U=2$ networks are marked by a linear dependence of the CLS set, such that the network reduces to $U=1$ networks in complete agreement with our analytical predictions. The experimentally studied sawtooth ST1 lattice is among the obtained $U=2$ set. 
We obtain a new and even simpler ST2 lattice with all hoppings being equal, which should be easily experimentally accessible.
Our results pave the way towards a complete description of FBs in networks with more bands and in higher dimensions, and towards
a full understanding of the topological properties of macroscopic degeneracies associated with FBs~\cite{peotta2015superfluidity,julku2016geometric,tovmaysan2016effective,read2016compactly,liang2016band}.
Note that in higher dimensions the Bravais lattice classification has to be used, which defines the number of intercell hopping matrices.
Further, we expect the shape of a CLS to become another classification property.

\begin{acknowledgments}
This work was supported by Project Code(IBS-R024-D1).
\end{acknowledgments}

\bibliography{flatband}

\begin{thebibliography}{33}%
\makeatletter
\providecommand \@ifxundefined [1]{%
 \@ifx{#1\undefined}
}%
\providecommand \@ifnum [1]{%
 \ifnum #1\expandafter \@firstoftwo
 \else \expandafter \@secondoftwo
 \fi
}%
\providecommand \@ifx [1]{%
 \ifx #1\expandafter \@firstoftwo
 \else \expandafter \@secondoftwo
 \fi
}%
\providecommand \natexlab [1]{#1}%
\providecommand \enquote  [1]{``#1''}%
\providecommand \bibnamefont  [1]{#1}%
\providecommand \bibfnamefont [1]{#1}%
\providecommand \citenamefont [1]{#1}%
\providecommand \href@noop [0]{\@secondoftwo}%
\providecommand \href [0]{\begingroup \@sanitize@url \@href}%
\providecommand \@href[1]{\@@startlink{#1}\@@href}%
\providecommand \@@href[1]{\endgroup#1\@@endlink}%
\providecommand \@sanitize@url [0]{\catcode `\\12\catcode `\$12\catcode
  `\&12\catcode `\#12\catcode `\^12\catcode `\_12\catcode `\%12\relax}%
\providecommand \@@startlink[1]{}%
\providecommand \@@endlink[0]{}%
\providecommand \url  [0]{\begingroup\@sanitize@url \@url }%
\providecommand \@url [1]{\endgroup\@href {#1}{\urlprefix }}%
\providecommand \urlprefix  [0]{URL }%
\providecommand \Eprint [0]{\href }%
\providecommand \doibase [0]{http://dx.doi.org/}%
\providecommand \selectlanguage [0]{\@gobble}%
\providecommand \bibinfo  [0]{\@secondoftwo}%
\providecommand \bibfield  [0]{\@secondoftwo}%
\providecommand \translation [1]{[#1]}%
\providecommand \BibitemOpen [0]{}%
\providecommand \bibitemStop [0]{}%
\providecommand \bibitemNoStop [0]{.\EOS\space}%
\providecommand \EOS [0]{\spacefactor3000\relax}%
\providecommand \BibitemShut  [1]{\csname bibitem#1\endcsname}%
\let\auto@bib@innerbib\@empty
\bibitem [{\citenamefont {Derzhko}\ \emph {et~al.}(2015)\citenamefont
  {Derzhko}, \citenamefont {Richter},\ and\ \citenamefont
  {Maksymenko}}]{derzhko2015strongly}%
  \BibitemOpen
  \bibfield  {author} {\bibinfo {author} {\bibfnamefont {Oleg}\ \bibnamefont
  {Derzhko}}, \bibinfo {author} {\bibfnamefont {Johannes}\ \bibnamefont
  {Richter}}, \ and\ \bibinfo {author} {\bibfnamefont {Mykola}\ \bibnamefont
  {Maksymenko}},\ }\bibfield  {title} {\enquote {\bibinfo {title} {Strongly
  correlated flat-band systems: The route from heisenberg spins to hubbard
  electrons},}\ }\href {\doibase 10.1142/S0217979215300078} {\bibfield
  {journal} {\bibinfo  {journal} {Int. J. Mod. Phys. B}\ }\textbf {\bibinfo
  {volume} {29}},\ \bibinfo {pages} {1530007} (\bibinfo {year}
  {2015})}\BibitemShut {NoStop}%
\bibitem [{\citenamefont {Mielke}(1991)}]{mielke1991ferromagnetism}%
  \BibitemOpen
  \bibfield  {author} {\bibinfo {author} {\bibfnamefont {A}~\bibnamefont
  {Mielke}},\ }\bibfield  {title} {\enquote {\bibinfo {title} {Ferromagnetism
  in the hubbard model on line graphs and further considerations},}\ }\href
  {http://stacks.iop.org/0305-4470/24/i=14/a=018} {\bibfield  {journal}
  {\bibinfo  {journal} {J. Phys. A: Math. Gen.}\ }\textbf {\bibinfo {volume}
  {24}},\ \bibinfo {pages} {3311} (\bibinfo {year} {1991})}\BibitemShut
  {NoStop}%
\bibitem [{\citenamefont {Tasaki}(1992)}]{tasaki1992ferromagnetism}%
  \BibitemOpen
  \bibfield  {author} {\bibinfo {author} {\bibfnamefont {Hal}\ \bibnamefont
  {Tasaki}},\ }\bibfield  {title} {\enquote {\bibinfo {title} {Ferromagnetism
  in the hubbard models with degenerate single-electron ground states},}\
  }\href {\doibase 10.1103/PhysRevLett.69.1608} {\bibfield  {journal} {\bibinfo
   {journal} {Phys. Rev. Lett.}\ }\textbf {\bibinfo {volume} {69}},\ \bibinfo
  {pages} {1608--1611} (\bibinfo {year} {1992})}\BibitemShut {NoStop}%
\bibitem [{\citenamefont {Derzhko}\ and\ \citenamefont
  {Richter}(2006)}]{derzhko2006universal}%
  \BibitemOpen
  \bibfield  {author} {\bibinfo {author} {\bibfnamefont {O.}~\bibnamefont
  {Derzhko}}\ and\ \bibinfo {author} {\bibfnamefont {J.}~\bibnamefont
  {Richter}},\ }\bibfield  {title} {\enquote {\bibinfo {title} {Universal
  low-temperature behavior of frustrated quantum antiferromagnets in the
  vicinity of the saturation field},}\ }\href {\doibase
  10.1140/epjb/e2006-00273-y} {\bibfield  {journal} {\bibinfo  {journal} {Eur.
  Phys. J. B - Cond. Mat. and Complex Sys.}\ }\textbf {\bibinfo {volume}
  {52}},\ \bibinfo {pages} {23--36} (\bibinfo {year} {2006})}\BibitemShut
  {NoStop}%
\bibitem [{\citenamefont {Derzhko}\ \emph {et~al.}(2010)\citenamefont
  {Derzhko}, \citenamefont {Richter}, \citenamefont {Honecker}, \citenamefont
  {Maksymenko},\ and\ \citenamefont {Moessner}}]{derzhko2010low}%
  \BibitemOpen
  \bibfield  {author} {\bibinfo {author} {\bibfnamefont {O.}~\bibnamefont
  {Derzhko}}, \bibinfo {author} {\bibfnamefont {J.}~\bibnamefont {Richter}},
  \bibinfo {author} {\bibfnamefont {A.}~\bibnamefont {Honecker}}, \bibinfo
  {author} {\bibfnamefont {M.}~\bibnamefont {Maksymenko}}, \ and\ \bibinfo
  {author} {\bibfnamefont {R.}~\bibnamefont {Moessner}},\ }\bibfield  {title}
  {\enquote {\bibinfo {title} {Low-temperature properties of the hubbard model
  on highly frustrated one-dimensional lattices},}\ }\href {\doibase
  10.1103/PhysRevB.81.014421} {\bibfield  {journal} {\bibinfo  {journal} {Phys.
  Rev. B}\ }\textbf {\bibinfo {volume} {81}},\ \bibinfo {pages} {014421}
  (\bibinfo {year} {2010})}\BibitemShut {NoStop}%
\bibitem [{\citenamefont {Hyrk\"as}\ \emph {et~al.}(2013)\citenamefont
  {Hyrk\"as}, \citenamefont {Apaja},\ and\ \citenamefont
  {Manninen}}]{hyrkas2013many}%
  \BibitemOpen
  \bibfield  {author} {\bibinfo {author} {\bibfnamefont {M.}~\bibnamefont
  {Hyrk\"as}}, \bibinfo {author} {\bibfnamefont {V.}~\bibnamefont {Apaja}}, \
  and\ \bibinfo {author} {\bibfnamefont {M.}~\bibnamefont {Manninen}},\
  }\bibfield  {title} {\enquote {\bibinfo {title} {Many-particle dynamics of
  bosons and fermions in quasi-one-dimensional flat-band lattices},}\ }\href
  {\doibase 10.1103/PhysRevA.87.023614} {\bibfield  {journal} {\bibinfo
  {journal} {Phys. Rev. A}\ }\textbf {\bibinfo {volume} {87}},\ \bibinfo
  {pages} {023614} (\bibinfo {year} {2013})}\BibitemShut {NoStop}%
\bibitem [{\citenamefont {Guzm{\'a}n-Silva}\ \emph {et~al.}(2014)\citenamefont
  {Guzm{\'a}n-Silva}, \citenamefont {Mej{\'i}�a-Cort{\'e}s}, \citenamefont
  {Bandres}, \citenamefont {Rechtsman}, \citenamefont {Weimann}, \citenamefont
  {Nolte}, \citenamefont {Segev}, \citenamefont {Szameit},\ and\ \citenamefont
  {Vicencio}}]{guzman2014experimental}%
  \BibitemOpen
  \bibfield  {author} {\bibinfo {author} {\bibfnamefont {D}~\bibnamefont
  {Guzm{\'a}n-Silva}}, \bibinfo {author} {\bibfnamefont {C}~\bibnamefont
  {Mej{\'i}�a-Cort{\'e}s}}, \bibinfo {author} {\bibfnamefont {M~A}\
  \bibnamefont {Bandres}}, \bibinfo {author} {\bibfnamefont {M~C}\ \bibnamefont
  {Rechtsman}}, \bibinfo {author} {\bibfnamefont {S}~\bibnamefont {Weimann}},
  \bibinfo {author} {\bibfnamefont {S}~\bibnamefont {Nolte}}, \bibinfo {author}
  {\bibfnamefont {M}~\bibnamefont {Segev}}, \bibinfo {author} {\bibfnamefont
  {A}~\bibnamefont {Szameit}}, \ and\ \bibinfo {author} {\bibfnamefont {R~A}\
  \bibnamefont {Vicencio}},\ }\bibfield  {title} {\enquote {\bibinfo {title}
  {Experimental observation of bulk and edge transport in photonic lieb
  lattices},}\ }\href {http://stacks.iop.org/1367-2630/16/i=6/a=063061}
  {\bibfield  {journal} {\bibinfo  {journal} {New J. Phys.}\ }\textbf {\bibinfo
  {volume} {16}},\ \bibinfo {pages} {063061} (\bibinfo {year}
  {2014})}\BibitemShut {NoStop}%
\bibitem [{\citenamefont {Vicencio}\ \emph {et~al.}(2015)\citenamefont
  {Vicencio}, \citenamefont {Cantillano}, \citenamefont {Morales-Inostroza},
  \citenamefont {Real}, \citenamefont {Mej\'{\i}a-Cort\'es}, \citenamefont
  {Weimann}, \citenamefont {Szameit},\ and\ \citenamefont
  {Molina}}]{vicencio2015observation}%
  \BibitemOpen
  \bibfield  {author} {\bibinfo {author} {\bibfnamefont {Rodrigo~A.}\
  \bibnamefont {Vicencio}}, \bibinfo {author} {\bibfnamefont {Camilo}\
  \bibnamefont {Cantillano}}, \bibinfo {author} {\bibfnamefont {Luis}\
  \bibnamefont {Morales-Inostroza}}, \bibinfo {author} {\bibfnamefont
  {Basti\'an}\ \bibnamefont {Real}}, \bibinfo {author} {\bibfnamefont
  {Cristian}\ \bibnamefont {Mej\'{\i}a-Cort\'es}}, \bibinfo {author}
  {\bibfnamefont {Steffen}\ \bibnamefont {Weimann}}, \bibinfo {author}
  {\bibfnamefont {Alexander}\ \bibnamefont {Szameit}}, \ and\ \bibinfo {author}
  {\bibfnamefont {Mario~I.}\ \bibnamefont {Molina}},\ }\bibfield  {title}
  {\enquote {\bibinfo {title} {Observation of localized states in lieb photonic
  lattices},}\ }\href {\doibase 10.1103/PhysRevLett.114.245503} {\bibfield
  {journal} {\bibinfo  {journal} {Phys. Rev. Lett.}\ }\textbf {\bibinfo
  {volume} {114}},\ \bibinfo {pages} {245503} (\bibinfo {year}
  {2015})}\BibitemShut {NoStop}%
\bibitem [{\citenamefont {Mukherjee}\ \emph {et~al.}(2015)\citenamefont
  {Mukherjee}, \citenamefont {Spracklen}, \citenamefont {Choudhury},
  \citenamefont {Goldman}, \citenamefont {\"Ohberg}, \citenamefont
  {Andersson},\ and\ \citenamefont {Thomson}}]{mukherjee2015observation}%
  \BibitemOpen
  \bibfield  {author} {\bibinfo {author} {\bibfnamefont {Sebabrata}\
  \bibnamefont {Mukherjee}}, \bibinfo {author} {\bibfnamefont {Alexander}\
  \bibnamefont {Spracklen}}, \bibinfo {author} {\bibfnamefont {Debaditya}\
  \bibnamefont {Choudhury}}, \bibinfo {author} {\bibfnamefont {Nathan}\
  \bibnamefont {Goldman}}, \bibinfo {author} {\bibfnamefont {Patrik}\
  \bibnamefont {\"Ohberg}}, \bibinfo {author} {\bibfnamefont {Erika}\
  \bibnamefont {Andersson}}, \ and\ \bibinfo {author} {\bibfnamefont
  {Robert~R.}\ \bibnamefont {Thomson}},\ }\bibfield  {title} {\enquote
  {\bibinfo {title} {Observation of a localized flat-band state in a photonic
  lieb lattice},}\ }\href {\doibase 10.1103/PhysRevLett.114.245504} {\bibfield
  {journal} {\bibinfo  {journal} {Phys. Rev. Lett.}\ }\textbf {\bibinfo
  {volume} {114}},\ \bibinfo {pages} {245504} (\bibinfo {year}
  {2015})}\BibitemShut {NoStop}%
\bibitem [{\citenamefont {Mukherjee}\ and\ \citenamefont
  {Thomson}(2015)}]{mukherjee2015observation1}%
  \BibitemOpen
  \bibfield  {author} {\bibinfo {author} {\bibfnamefont {Sebabrata}\
  \bibnamefont {Mukherjee}}\ and\ \bibinfo {author} {\bibfnamefont {Robert~R.}\
  \bibnamefont {Thomson}},\ }\bibfield  {title} {\enquote {\bibinfo {title}
  {Observation of localized flat-band modes in a quasi-one-dimensional photonic
  rhombic lattice},}\ }\href {\doibase 10.1364/OL.40.005443} {\bibfield
  {journal} {\bibinfo  {journal} {Opt. Lett.}\ }\textbf {\bibinfo {volume}
  {40}},\ \bibinfo {pages} {5443--5446} (\bibinfo {year} {2015})}\BibitemShut
  {NoStop}%
\bibitem [{\citenamefont {Weimann}\ \emph {et~al.}(2016)\citenamefont
  {Weimann}, \citenamefont {Morales-Inostroza}, \citenamefont {Real},
  \citenamefont {Cantillano}, \citenamefont {Szameit},\ and\ \citenamefont
  {Vicencio}}]{weimann2016transport}%
  \BibitemOpen
  \bibfield  {author} {\bibinfo {author} {\bibfnamefont {Steffen}\ \bibnamefont
  {Weimann}}, \bibinfo {author} {\bibfnamefont {Luis}\ \bibnamefont
  {Morales-Inostroza}}, \bibinfo {author} {\bibfnamefont {Basti\'{a}n}\
  \bibnamefont {Real}}, \bibinfo {author} {\bibfnamefont {Camilo}\ \bibnamefont
  {Cantillano}}, \bibinfo {author} {\bibfnamefont {Alexander}\ \bibnamefont
  {Szameit}}, \ and\ \bibinfo {author} {\bibfnamefont {Rodrigo~A.}\
  \bibnamefont {Vicencio}},\ }\bibfield  {title} {\enquote {\bibinfo {title}
  {Transport in sawtooth photonic lattices},}\ }\href {\doibase
  10.1364/OL.41.002414} {\bibfield  {journal} {\bibinfo  {journal} {Opt.
  Lett.}\ }\textbf {\bibinfo {volume} {41}},\ \bibinfo {pages} {2414--2417}
  (\bibinfo {year} {2016})}\BibitemShut {NoStop}%
\bibitem [{\citenamefont {Xia}\ \emph {et~al.}(2016)\citenamefont {Xia},
  \citenamefont {Hu}, \citenamefont {Song}, \citenamefont {Zong}, \citenamefont
  {Tang},\ and\ \citenamefont {Chen}}]{xia2016demonstration}%
  \BibitemOpen
  \bibfield  {author} {\bibinfo {author} {\bibfnamefont {Shiqiang}\
  \bibnamefont {Xia}}, \bibinfo {author} {\bibfnamefont {Yi}~\bibnamefont
  {Hu}}, \bibinfo {author} {\bibfnamefont {Daohong}\ \bibnamefont {Song}},
  \bibinfo {author} {\bibfnamefont {Yuanyuan}\ \bibnamefont {Zong}}, \bibinfo
  {author} {\bibfnamefont {Liqin}\ \bibnamefont {Tang}}, \ and\ \bibinfo
  {author} {\bibfnamefont {Zhigang}\ \bibnamefont {Chen}},\ }\bibfield  {title}
  {\enquote {\bibinfo {title} {Demonstration of flat-band image transmission in
  optically induced lieb photonic lattices},}\ }\href {\doibase
  10.1364/OL.41.001435} {\bibfield  {journal} {\bibinfo  {journal} {Opt.
  Lett.}\ }\textbf {\bibinfo {volume} {41}},\ \bibinfo {pages} {1435--1438}
  (\bibinfo {year} {2016})}\BibitemShut {NoStop}%
\bibitem [{\citenamefont {Masumoto}\ \emph {et~al.}(2012)\citenamefont
  {Masumoto}, \citenamefont {Kim}, \citenamefont {Byrnes}, \citenamefont
  {Kusudo}, \citenamefont {L{\"o}ffler}, \citenamefont {H{\"o}fling},
  \citenamefont {Forchel},\ and\ \citenamefont
  {Yamamoto}}]{masumoto2012exciton}%
  \BibitemOpen
  \bibfield  {author} {\bibinfo {author} {\bibfnamefont {Naoyuki}\ \bibnamefont
  {Masumoto}}, \bibinfo {author} {\bibfnamefont {Na~Young}\ \bibnamefont
  {Kim}}, \bibinfo {author} {\bibfnamefont {Tim}\ \bibnamefont {Byrnes}},
  \bibinfo {author} {\bibfnamefont {Kenichiro}\ \bibnamefont {Kusudo}},
  \bibinfo {author} {\bibfnamefont {Andreas}\ \bibnamefont {L{\"o}ffler}},
  \bibinfo {author} {\bibfnamefont {Sven}\ \bibnamefont {H{\"o}fling}},
  \bibinfo {author} {\bibfnamefont {Alfred}\ \bibnamefont {Forchel}}, \ and\
  \bibinfo {author} {\bibfnamefont {Yoshihisa}\ \bibnamefont {Yamamoto}},\
  }\bibfield  {title} {\enquote {\bibinfo {title} {Exciton--polariton
  condensates with flat bands in a two-dimensional kagome lattice},}\ }\href
  {http://stacks.iop.org/1367-2630/14/i=6/a=065002} {\bibfield  {journal}
  {\bibinfo  {journal} {New J. Phys.}\ }\textbf {\bibinfo {volume} {14}},\
  \bibinfo {pages} {065002} (\bibinfo {year} {2012})}\BibitemShut {NoStop}%
\bibitem [{\citenamefont {Baboux}\ \emph {et~al.}(2016)\citenamefont {Baboux},
  \citenamefont {Ge}, \citenamefont {Jacqmin}, \citenamefont {Biondi},
  \citenamefont {Galopin}, \citenamefont {Lema\^{\i}tre}, \citenamefont
  {Le~Gratiet}, \citenamefont {Sagnes}, \citenamefont {Schmidt}, \citenamefont
  {T\"ureci}, \citenamefont {Amo},\ and\ \citenamefont
  {Bloch}}]{baboux2016bosonic}%
  \BibitemOpen
  \bibfield  {author} {\bibinfo {author} {\bibfnamefont {F.}~\bibnamefont
  {Baboux}}, \bibinfo {author} {\bibfnamefont {L.}~\bibnamefont {Ge}}, \bibinfo
  {author} {\bibfnamefont {T.}~\bibnamefont {Jacqmin}}, \bibinfo {author}
  {\bibfnamefont {M.}~\bibnamefont {Biondi}}, \bibinfo {author} {\bibfnamefont
  {E.}~\bibnamefont {Galopin}}, \bibinfo {author} {\bibfnamefont
  {A.}~\bibnamefont {Lema\^{\i}tre}}, \bibinfo {author} {\bibfnamefont
  {L.}~\bibnamefont {Le~Gratiet}}, \bibinfo {author} {\bibfnamefont
  {I.}~\bibnamefont {Sagnes}}, \bibinfo {author} {\bibfnamefont
  {S.}~\bibnamefont {Schmidt}}, \bibinfo {author} {\bibfnamefont {H.~E.}\
  \bibnamefont {T\"ureci}}, \bibinfo {author} {\bibfnamefont {A.}~\bibnamefont
  {Amo}}, \ and\ \bibinfo {author} {\bibfnamefont {J.}~\bibnamefont {Bloch}},\
  }\bibfield  {title} {\enquote {\bibinfo {title} {Bosonic condensation and
  disorder-induced localization in a flat band},}\ }\href {\doibase
  10.1103/PhysRevLett.116.066402} {\bibfield  {journal} {\bibinfo  {journal}
  {Phys. Rev. Lett.}\ }\textbf {\bibinfo {volume} {116}},\ \bibinfo {pages}
  {066402} (\bibinfo {year} {2016})}\BibitemShut {NoStop}%
\bibitem [{\citenamefont {Taie}\ \emph {et~al.}(2015)\citenamefont {Taie},
  \citenamefont {Ozawa}, \citenamefont {Ichinose}, \citenamefont {Nishio},
  \citenamefont {Nakajima},\ and\ \citenamefont
  {Takahashi}}]{taie2015coherent}%
  \BibitemOpen
  \bibfield  {author} {\bibinfo {author} {\bibfnamefont {Shintaro}\
  \bibnamefont {Taie}}, \bibinfo {author} {\bibfnamefont {Hideki}\ \bibnamefont
  {Ozawa}}, \bibinfo {author} {\bibfnamefont {Tomohiro}\ \bibnamefont
  {Ichinose}}, \bibinfo {author} {\bibfnamefont {Takuei}\ \bibnamefont
  {Nishio}}, \bibinfo {author} {\bibfnamefont {Shuta}\ \bibnamefont
  {Nakajima}}, \ and\ \bibinfo {author} {\bibfnamefont {Yoshiro}\ \bibnamefont
  {Takahashi}},\ }\bibfield  {title} {\enquote {\bibinfo {title} {Coherent
  driving and freezing of bosonic matter wave in an optical lieb lattice},}\
  }\href {\doibase 10.1126/sciadv.1500854} {\bibfield  {journal} {\bibinfo
  {journal} {Sci. Adv.}\ }\textbf {\bibinfo {volume} {1}} (\bibinfo {year}
  {2015}),\ 10.1126/sciadv.1500854}\BibitemShut {NoStop}%
\bibitem [{\citenamefont {Jo}\ \emph {et~al.}(2012)\citenamefont {Jo},
  \citenamefont {Guzman}, \citenamefont {Thomas}, \citenamefont {Hosur},
  \citenamefont {Vishwanath},\ and\ \citenamefont
  {Stamper-Kurn}}]{jo2012ultracold}%
  \BibitemOpen
  \bibfield  {author} {\bibinfo {author} {\bibfnamefont {Gyu-Boong}\
  \bibnamefont {Jo}}, \bibinfo {author} {\bibfnamefont {Jennie}\ \bibnamefont
  {Guzman}}, \bibinfo {author} {\bibfnamefont {Claire~K.}\ \bibnamefont
  {Thomas}}, \bibinfo {author} {\bibfnamefont {Pavan}\ \bibnamefont {Hosur}},
  \bibinfo {author} {\bibfnamefont {Ashvin}\ \bibnamefont {Vishwanath}}, \ and\
  \bibinfo {author} {\bibfnamefont {Dan~M.}\ \bibnamefont {Stamper-Kurn}},\
  }\bibfield  {title} {\enquote {\bibinfo {title} {Ultracold atoms in a tunable
  optical kagome lattice},}\ }\href {\doibase 10.1103/PhysRevLett.108.045305}
  {\bibfield  {journal} {\bibinfo  {journal} {Phys. Rev. Lett.}\ }\textbf
  {\bibinfo {volume} {108}},\ \bibinfo {pages} {045305} (\bibinfo {year}
  {2012})}\BibitemShut {NoStop}%
\bibitem [{\citenamefont {Flach}\ \emph {et~al.}(2014)\citenamefont {Flach},
  \citenamefont {Leykam}, \citenamefont {Bodyfelt}, \citenamefont {Matthies},\
  and\ \citenamefont {Desyatnikov}}]{flach2014detangling}%
  \BibitemOpen
  \bibfield  {author} {\bibinfo {author} {\bibfnamefont {Sergej}\ \bibnamefont
  {Flach}}, \bibinfo {author} {\bibfnamefont {Daniel}\ \bibnamefont {Leykam}},
  \bibinfo {author} {\bibfnamefont {Joshua~D.}\ \bibnamefont {Bodyfelt}},
  \bibinfo {author} {\bibfnamefont {Peter}\ \bibnamefont {Matthies}}, \ and\
  \bibinfo {author} {\bibfnamefont {Anton~S.}\ \bibnamefont {Desyatnikov}},\
  }\bibfield  {title} {\enquote {\bibinfo {title} {Detangling flat bands into
  fano lattices},}\ }\href {http://stacks.iop.org/0295-5075/105/i=3/a=30001}
  {\bibfield  {journal} {\bibinfo  {journal} {EPL (Europhysics Letters)}\
  }\textbf {\bibinfo {volume} {105}},\ \bibinfo {pages} {30001} (\bibinfo
  {year} {2014})}\BibitemShut {NoStop}%
\bibitem [{\citenamefont {Sachdev}(2007)}]{sachdev2007quantum}%
  \BibitemOpen
  \bibfield  {author} {\bibinfo {author} {\bibfnamefont {Subir}\ \bibnamefont
  {Sachdev}},\ }\href {\doibase 10.1002/9780470022184.hmm108} {\emph {\bibinfo
  {title} {Handbook of Magnetism and Advanced Magnetic Materials}}}\ (\bibinfo
  {publisher} {John Wiley and Sons, Ltd},\ \bibinfo {year} {2007})\BibitemShut
  {NoStop}%
\bibitem [{\citenamefont {Leykam}\ \emph {et~al.}(2016)\citenamefont {Leykam},
  \citenamefont {Bodyfelt}, \citenamefont {Desyatnikov},\ and\ \citenamefont
  {Flach}}]{leykam2016localisation}%
  \BibitemOpen
  \bibfield  {author} {\bibinfo {author} {\bibfnamefont {Daniel}\ \bibnamefont
  {Leykam}}, \bibinfo {author} {\bibfnamefont {Joshua~D.}\ \bibnamefont
  {Bodyfelt}}, \bibinfo {author} {\bibfnamefont {Anton~S.}\ \bibnamefont
  {Desyatnikov}}, \ and\ \bibinfo {author} {\bibfnamefont {Sergej}\
  \bibnamefont {Flach}},\ }\href@noop {} {\enquote {\bibinfo {title}
  {Localization of weakly disordered flat band states},}\ } (\bibinfo {year}
  {2016}),\ \Eprint {http://arxiv.org/abs/1601.03784} {arXiv:1601.03784
  [cond-mat]} \BibitemShut {NoStop}%
\bibitem [{\citenamefont {Bodyfelt}\ \emph {et~al.}(2014)\citenamefont
  {Bodyfelt}, \citenamefont {Leykam}, \citenamefont {Danieli}, \citenamefont
  {Yu},\ and\ \citenamefont {Flach}}]{bodyfelt2014flatbands}%
  \BibitemOpen
  \bibfield  {author} {\bibinfo {author} {\bibfnamefont {Joshua~D.}\
  \bibnamefont {Bodyfelt}}, \bibinfo {author} {\bibfnamefont {Daniel}\
  \bibnamefont {Leykam}}, \bibinfo {author} {\bibfnamefont {Carlo}\
  \bibnamefont {Danieli}}, \bibinfo {author} {\bibfnamefont {Xiaoquan}\
  \bibnamefont {Yu}}, \ and\ \bibinfo {author} {\bibfnamefont {Sergej}\
  \bibnamefont {Flach}},\ }\bibfield  {title} {\enquote {\bibinfo {title}
  {Flatbands under correlated perturbations},}\ }\href {\doibase
  10.1103/PhysRevLett.113.236403} {\bibfield  {journal} {\bibinfo  {journal}
  {Phys. Rev. Lett.}\ }\textbf {\bibinfo {volume} {113}},\ \bibinfo {pages}
  {236403} (\bibinfo {year} {2014})}\BibitemShut {NoStop}%
\bibitem [{\citenamefont {Danieli}\ \emph {et~al.}(2015)\citenamefont
  {Danieli}, \citenamefont {Bodyfelt},\ and\ \citenamefont
  {Flach}}]{danieli2015flatband}%
  \BibitemOpen
  \bibfield  {author} {\bibinfo {author} {\bibfnamefont {Carlo}\ \bibnamefont
  {Danieli}}, \bibinfo {author} {\bibfnamefont {Joshua~D.}\ \bibnamefont
  {Bodyfelt}}, \ and\ \bibinfo {author} {\bibfnamefont {Sergej}\ \bibnamefont
  {Flach}},\ }\bibfield  {title} {\enquote {\bibinfo {title} {Flat-band
  engineering of mobility edges},}\ }\href {\doibase
  10.1103/PhysRevB.91.235134} {\bibfield  {journal} {\bibinfo  {journal} {Phys.
  Rev. B}\ }\textbf {\bibinfo {volume} {91}},\ \bibinfo {pages} {235134}
  (\bibinfo {year} {2015})}\BibitemShut {NoStop}%
\bibitem [{\citenamefont {Khomeriki}\ and\ \citenamefont
  {Flach}(2016)}]{khomeriki2016landau}%
  \BibitemOpen
  \bibfield  {author} {\bibinfo {author} {\bibfnamefont {Ramaz}\ \bibnamefont
  {Khomeriki}}\ and\ \bibinfo {author} {\bibfnamefont {Sergej}\ \bibnamefont
  {Flach}},\ }\bibfield  {title} {\enquote {\bibinfo {title} {Landau-zener
  bloch oscillations with perturbed flat bands},}\ }\href {\doibase
  10.1103/PhysRevLett.116.245301} {\bibfield  {journal} {\bibinfo  {journal}
  {Phys. Rev. Lett.}\ }\textbf {\bibinfo {volume} {116}},\ \bibinfo {pages}
  {245301} (\bibinfo {year} {2016})}\BibitemShut {NoStop}%
\bibitem [{\citenamefont {Dias}\ and\ \citenamefont
  {D.}(2015)}]{dias2015origami}%
  \BibitemOpen
  \bibfield  {author} {\bibinfo {author} {\bibfnamefont {R.~G.}\ \bibnamefont
  {Dias}}\ and\ \bibinfo {author} {\bibfnamefont {Gouveia~J.}\ \bibnamefont
  {D.}},\ }\bibfield  {title} {\enquote {\bibinfo {title} {Origami rules for
  the construction of localized eigenstates of the hubbard model in decorated
  lattices},}\ }\href {\doibase 10.1038/srep16852} {\bibfield  {journal}
  {\bibinfo  {journal} {Sci. Rep.}\ }\textbf {\bibinfo {volume} {5}},\ \bibinfo
  {pages} {16852} (\bibinfo {year} {2015})}\BibitemShut {NoStop}%
\bibitem [{\citenamefont {Morales-Inostroza}\ and\ \citenamefont
  {Vicencio}(2016)}]{morales2016simple}%
  \BibitemOpen
  \bibfield  {author} {\bibinfo {author} {\bibfnamefont {Luis}\ \bibnamefont
  {Morales-Inostroza}}\ and\ \bibinfo {author} {\bibfnamefont {Rodrigo~A.}\
  \bibnamefont {Vicencio}},\ }\bibfield  {title} {\enquote {\bibinfo {title}
  {Simple method to construct flat-band lattices},}\ }\href {\doibase
  10.1103/PhysRevA.94.043831} {\bibfield  {journal} {\bibinfo  {journal} {Phys.
  Rev. A}\ }\textbf {\bibinfo {volume} {94}},\ \bibinfo {pages} {043831}
  (\bibinfo {year} {2016})}\BibitemShut {NoStop}%
\bibitem [{\citenamefont {Vidal}\ \emph {et~al.}(1998)\citenamefont {Vidal},
  \citenamefont {Mosseri},\ and\ \citenamefont {Dou\ifmmode~\mbox{\c{c}}\else
  \c{c}\fi{}ot}}]{vidal1998aharonov}%
  \BibitemOpen
  \bibfield  {author} {\bibinfo {author} {\bibfnamefont {Julien}\ \bibnamefont
  {Vidal}}, \bibinfo {author} {\bibfnamefont {R\'emy}\ \bibnamefont {Mosseri}},
  \ and\ \bibinfo {author} {\bibfnamefont {Benoit}\ \bibnamefont
  {Dou\ifmmode~\mbox{\c{c}}\else \c{c}\fi{}ot}},\ }\bibfield  {title} {\enquote
  {\bibinfo {title} {Aharonov-bohm cages in two-dimensional structures},}\
  }\href {\doibase 10.1103/PhysRevLett.81.5888} {\bibfield  {journal} {\bibinfo
   {journal} {Phys. Rev. Lett.}\ }\textbf {\bibinfo {volume} {81}},\ \bibinfo
  {pages} {5888--5891} (\bibinfo {year} {1998})}\BibitemShut {NoStop}%
\bibitem [{Note1()}]{Note1}%
  \BibitemOpen
  \bibinfo {note} {Note that $H_0$ is Hermitian, while $H_m$ with $m\not =0$
  are not in general.}\BibitemShut {Stop}%
\bibitem [{sup()}]{supp}%
  \BibitemOpen
  \bibfield  {title} {\enquote {\bibinfo {title} {See supplemental material at
  url to be defined.}}\ }\href@noop {} {\ }\BibitemShut {NoStop}%
\bibitem [{Note2()}]{Note2}%
  \BibitemOpen
  \bibinfo {note} {This conclusion can also be verified e.g. using a band
  structure calculation.}\BibitemShut {Stop}%
\bibitem [{\citenamefont {Peotta}\ and\ \citenamefont
  {T{\"{o}}rm{\"{a}}}(2015)}]{peotta2015superfluidity}%
  \BibitemOpen
  \bibfield  {author} {\bibinfo {author} {\bibfnamefont {Sebastiano}\
  \bibnamefont {Peotta}}\ and\ \bibinfo {author} {\bibfnamefont {P{\"{a}}ivi}\
  \bibnamefont {T{\"{o}}rm{\"{a}}}},\ }\bibfield  {title} {\enquote {\bibinfo
  {title} {Superfluidity in topologically nontrivial flat bands},}\ }\href
  {http://dx.doi.org/10.1038/ncomms9944 http://10.0.4.14/ncomms9944
  http://www.nature.com/articles/ncomms9944{\#}supplementary-information}
  {\bibfield  {journal} {\bibinfo  {journal} {Nat. Comm.}\ }\textbf {\bibinfo
  {volume} {6}},\ \bibinfo {pages} {8944} (\bibinfo {year} {2015})}\BibitemShut
  {NoStop}%
\bibitem [{\citenamefont {Julku}\ \emph {et~al.}(2016)\citenamefont {Julku},
  \citenamefont {Peotta}, \citenamefont {Vanhala}, \citenamefont {Kim},\ and\
  \citenamefont {T\"orm\"a}}]{julku2016geometric}%
  \BibitemOpen
  \bibfield  {author} {\bibinfo {author} {\bibfnamefont {Aleksi}\ \bibnamefont
  {Julku}}, \bibinfo {author} {\bibfnamefont {Sebastiano}\ \bibnamefont
  {Peotta}}, \bibinfo {author} {\bibfnamefont {Tuomas~I.}\ \bibnamefont
  {Vanhala}}, \bibinfo {author} {\bibfnamefont {Dong-Hee}\ \bibnamefont {Kim}},
  \ and\ \bibinfo {author} {\bibfnamefont {P\"aivi}\ \bibnamefont
  {T\"orm\"a}},\ }\bibfield  {title} {\enquote {\bibinfo {title} {Geometric
  origin of superfluidity in the lieb-lattice flat band},}\ }\href {\doibase
  10.1103/PhysRevLett.117.045303} {\bibfield  {journal} {\bibinfo  {journal}
  {Phys. Rev. Lett.}\ }\textbf {\bibinfo {volume} {117}},\ \bibinfo {pages}
  {045303} (\bibinfo {year} {2016})}\BibitemShut {NoStop}%
\bibitem [{\citenamefont {Tovmasyan}\ \emph {et~al.}(2016)\citenamefont
  {Tovmasyan}, \citenamefont {Peotta}, \citenamefont {T{\"o}rm{\"a}},\ and\
  \citenamefont {Huber}}]{tovmaysan2016effective}%
  \BibitemOpen
  \bibfield  {author} {\bibinfo {author} {\bibfnamefont {M.}~\bibnamefont
  {Tovmasyan}}, \bibinfo {author} {\bibfnamefont {S.}~\bibnamefont {Peotta}},
  \bibinfo {author} {\bibfnamefont {P.}~\bibnamefont {T{\"o}rm{\"a}}}, \ and\
  \bibinfo {author} {\bibfnamefont {S.~D.}\ \bibnamefont {Huber}},\ }\href@noop
  {} {\enquote {\bibinfo {title} {Effective theory and emergent $su(2)$
  symmetry in the flat bands of attractive hubbard models},}\ } (\bibinfo
  {year} {2016}),\ \Eprint {http://arxiv.org/abs/1608.00976} {arXiv:1608.00976
  [cond-mat.str-el]} \BibitemShut {NoStop}%
\bibitem [{\citenamefont {Read}(2016)}]{read2016compactly}%
  \BibitemOpen
  \bibfield  {author} {\bibinfo {author} {\bibfnamefont {N.}~\bibnamefont
  {Read}},\ }\href@noop {} {\enquote {\bibinfo {title} {Compactly-supported
  wannier functions and algebraic $k$-theory},}\ } (\bibinfo {year} {2016}),\
  \Eprint {http://arxiv.org/abs/1608.04696} {arXiv:1608.04696
  [cond-mat.mes-hall]} \BibitemShut {NoStop}%
\bibitem [{\citenamefont {Liang}\ \emph {et~al.}(2016)\citenamefont {Liang},
  \citenamefont {Vanhala}, \citenamefont {Peotta}, \citenamefont {Siro},
  \citenamefont {Harju},\ and\ \citenamefont {T{\"o}rm{\"a}}}]{liang2016band}%
  \BibitemOpen
  \bibfield  {author} {\bibinfo {author} {\bibfnamefont {L.}~\bibnamefont
  {Liang}}, \bibinfo {author} {\bibfnamefont {T.~I.}\ \bibnamefont {Vanhala}},
  \bibinfo {author} {\bibfnamefont {S.}~\bibnamefont {Peotta}}, \bibinfo
  {author} {\bibfnamefont {T.}~\bibnamefont {Siro}}, \bibinfo {author}
  {\bibfnamefont {A.}~\bibnamefont {Harju}}, \ and\ \bibinfo {author}
  {\bibfnamefont {P.}~\bibnamefont {T{\"o}rm{\"a}}},\ }\href@noop {} {\enquote
  {\bibinfo {title} {Band geometry, berry curvature and superfluid weight},}\ }
  (\bibinfo {year} {2016}),\ \Eprint {http://arxiv.org/abs/1610.01803}
  {arXiv:1610.01803 [cond-mat.supr-con]} \BibitemShut {NoStop}%
\end{thebibliography}%

\setcounter{equation}{0}

\widetext

\newpage
{\bf Supplemental Material for ``Flatband generators with compact localized
states in one dimension''}
\\
\\
This Supplemental Material details the derivations of 1) the condition of linear independence of CLS, the reduction of linearly dependent and orthogonal $U=2$ CLS to $U=1$ CLS, the computation of the corresponding Bloch polarization vectors for linearly independent CLS; 2) the flatband generator for $U=2$ and the corresponding band structure. 3) the generalized sawtooth chain.

\section{On the linear independence of CLS and their corresponding Bloch polarization vectors}

\subsection*{Sufficient condition for a linearly independent set of CLS}

Consider a linearly dependent set of CLS $\vec{\Psi}_i$ such that
\begin{equation}
    \label{eq:lin-dep-cls}
    \sum_{j=-\infty}^\infty\alpha_j\vec{\Psi}_j = 0,\ \ j\in\mathbb{Z}
\end{equation}
where $\vec{\Psi}_j=\left(\dots0,0,\vec{\psi}_1,\vec{\psi}_2,\dots,\vec{\psi}_U,0,0,\dots\right)$, $\vec{\psi}_{1...U}$ are vectors with $\nu$ components, and $\vec{\psi}_1$ is located in the $j$th unit cell. A necessary condition for that is 
\begin{equation}
    \sum_{l=1}^{U}\alpha_{j+l}\vec{\psi}_l = 0\;,
\end{equation}
and thus the set $\{\vec{\psi}_l\}_{l=1,2,...,U}$ has to be linearly dependent.

If therefore $\{\vec{\psi}_l\}_{l=1,2,...,U}$ is a linearly independent set, the CLS set is linearly independent as well. This result is true for any values of $\nu,m_c,U$ in one dimension.

Since the dimension of a linearly independent CLS set is equal to the number of unit cells on the lattice, it will span the entire Hilbert space of the flat band.

\subsection*{Linear dependence for $m_c=1$ and $U=2$}

Consider a CLS of $U=2$ class $\vec{\psi}=\left(\vec{\psi}_1,\vec{\psi}_2\right)$,
with $m_c=1$. Assume that the two components $\vec{\psi}_1,\vec{\psi}_2$
are linearly dependent such that $\vec{\psi}_1=a\vec{\psi}_2$.
Since $\left(\vec{\psi}_1,\vec{\psi}_2\right)$ is a CLS, it follows
\begin{equation}
    \begin{split}
        H_1\vec{\psi}_1 & =0\;,\\
        H_1^{\dagger}\vec{\psi}_2 & =0\;.
    \end{split}
    \label{eq:u2-cls-con}
\end{equation}
This yields 
\begin{equation}
aH_1\vec{\psi}_2 = 0\;,\;H_1^\dagger\vec{\psi}_2 = 0\;\;\textrm{or}\;\;H_1\vec{\psi}_1 = 0\;,\;\frac{1}{a}H_1^\dagger\vec{\psi}_1 = 0\;.
\end{equation}
Thus $\vec{\psi}_1,\vec{\psi}_2$ are left and right eigenvectors of $H_1$ at the same time, therefore either $\vec{\psi}_1$ or $\vec{\psi}_2$ serves as the only component of a CLS of class $U=1$.

Interestingly a similar (but more lengthy) proof can be obtained for $\nu=2, m_c=1,\ U=3$. Given a CLS $\vec{\psi}=\left(\vec{\psi}_1,\vec{\psi}_2,\vec{\psi}_3\right)$, it can be shown that the linear dependence of $\{\vec{\psi}_1,\vec{\psi}_2,\vec{\psi}_3\}$ implies that the flat band is of class $U\leq2$. Once $\nu>2$ linear dependence is only a necessary, but not a sufficient condition.

\subsection*{Orthogonality for $\nu=2$ and $U=2$ }

Consider $U=2$ with $\vec{\psi}_1\perp\vec{\psi}_2$. Then a suitable rotation of the basis in each unit cell will result in $\vec{\psi}_1=(1,0)$ and $\vec{\psi}_2=(0,1)$. This seeming $U=2$ case can be reduced to $U=1$ by redefining the unit cell. Indeed after the above rotation we may denote each site in a unit cell by $a_l$ and $b_l$. Then a CLS is given by $a_l=\delta_{l,l_{0}}$ and $b_l=\delta_{l,l_{0}+1}$ (up to prefactors and renormalization factors). Redefining the unit cell using $\tilde{a}_l=a_l$ and $\tilde{b}_l=b_{l+1}$ turns the above CLS into class $U=1$. Therefore we can conclude, that for $m_c=1$, $\nu=2$ and $U=2$ $\vec{\psi}_1$ and $\vec{\psi}_2$ must be neither parallel (linearly dependent) nor orthogonal in order for the flat band to be not reducible to $U=1$.

\subsection*{From CLS to Bloch polarization vectors}

Transforming the original Hamiltonian into Bloch representation yields the Bloch Hamiltonian of rank $\nu$: 
\begin{equation}
    H(k) = \sum_{-m_c}^{m_c}H_m\mathrm{e}^{ikm}\;.
\end{equation}
The eigenvectors of $H(k)$ are the polarization vectors $\vec{u}(k)=\vec{u}(k+2\pi)$ which lead to the Bloch eigenstates of the original Hamiltonian with component $\vec{u}(k)\mathrm{e}^{ikl}$ on each unit cell. The set of CLS is given by $\vec{\Psi}_l=\left(\dots0,0,\vec{\psi}_1,\vec{\psi}_2,\dots,\vec{\psi}_U,0,0,\dots\right)$, and $\vec{\psi}_1$ is located in the $l$th unit cell. Since all CLS of a flat band share the same eigenenergy we can construct a new Bloch eigenstate (up to normalization) by computing 
\begin{equation}
    \sum_{l=-\infty}^{\infty}\mathrm{e}^{ikl}\vec{\Psi}_l\;.
\end{equation}
It follows that 
\begin{equation}
    \vec{u}(k)\sim\sum_{l=1}^U\vec{\psi}_l e^{i(U-1)k}\;.
\end{equation}
The inverse is also true: if a polarization vector for a certain band can be expressed as
\begin{equation}
    \vec{u}(k)=N(k)\sum_{l=1}^U\vec{\psi}_l e^{i(U-1)k},
\end{equation}
where $N(k)$ is a common prefactor, then the CLS state of class U is given by $\vec{\Psi}_l$ as defined above. Note however, that this $U$ need not be the smallest possible one, i.e. the eigenstate can be decomposed into even smaller CLS.

\section{Generator and band structure for $U=2$ FB networks}

For $\nu=2,\ m_c=1,\ U=2$ case we solve the following equations
\begin{align}
    \label{eq:eigenvalue_eq1}
    H_0\vec{\psi}_1 + H_1\vec{\psi}_2 & = & \efb\vec{\psi}_1\;,\\
    \label{eq:eigenvalue_eq2}
    H_1^\dagger\vec{\psi}_1 + H_0\vec{\psi}_2 & = & \efb\vec{\psi}_2\;,\\
    \label{eq:u2-cls-con1}
    H_1\vec{\psi}_1 & = & 0\;, \\
    \label{eq:u2-cls-con2}
    H_1^\dagger\vec{\psi}_2 & = & 0 \;.
\end{align}
We can always diagonalize $H_0$, and gauge and rescale the full Hamiltonian to obtain
\begin{equation}
    \label{eq:canonical-H0}
    H_0 = \left(\begin{array}{cc}
        0 & 0\\
        0 & 1
    \end{array}\right) \;.
\end{equation}
For non-singular $\Lambda=\efb-H_0$, we find $\vec{\psi}_2$ from~\eqref{eq:eigenvalue_eq2} and insert into~\eqref{eq:eigenvalue_eq1} to get 
\begin{eqnarray}
    \label{eq:new_eigenvalue_eq1}
    \Lambda^{-1}H_1\Lambda^{-1}H_1^\dagger\vec{\psi}_1 & = & \vec{\psi}_1\;, \\
    \Lambda^{-1}H_1^{\dagger}\vec{\psi}_1 & = & \vec{\psi}_2 \;, 
\end{eqnarray}
where $\Lambda^{-1}=\frac{1}{\efb-H_0}$. Similarly we have 
\begin{eqnarray}
    \label{eq:new_eigenvalue_eq2}
    \Lambda^{-1}H_1^\dagger\Lambda^{-1}H_1\vec{\psi}_2 & = & \vec{\psi}_2\;, \\
    \Lambda^{-1}H_1\vec{\psi}_2 & = & \vec{\psi}_1 \;.
\end{eqnarray}

\subsection*{Real $H_1$}

Consider all elements of $H_1$ to be real valued. Equations~\eqref{eq:u2-cls-con1} and~\eqref{eq:u2-cls-con2} allow to redefine $H_1$ in terms of unit vectors $\vert\varphi\rangle$ and $\vert\theta\rangle$ which are the left and right eigenvectors of the non-zero eigenvalue of $H_1$, and which are orthogonal to $\vec{\psi}_1$ and $\vec{\psi}_2$:
\begin{equation}
    \begin{aligned}H_1 & =\alpha\vert\theta\rangle\langle\varphi\vert \;, \\
        \vert\varphi\rangle & =\left(\begin{array}{c}
        \cos\varphi\\
        \sin\varphi
    \end{array}\right) \;, \\
    \vert\theta\rangle & =\left(\begin{array}{c}
        \cos\theta\\
        \sin\theta
    \end{array}\right) \;, 
    \end{aligned}
    \label{eq:redef_H1}
\end{equation}
where the scalar products
\begin{eqnarray}
    \label{eq:redef_cls_con1}
    \langle\vec{\psi}_1\vert\varphi\rangle & = & 0\;, \\
    \label{eq:redef_cls_con2}
    \langle\vec{\psi}_2\vert\theta\rangle & = & 0 \;.
\end{eqnarray}
Using these definitions and solving equations~\eqref{eq:new_eigenvalue_eq1} and~\eqref{eq:u2-cls-con1} we obtain
\begin{eqnarray}
    \label{eq:eigen_value}
    \efb & = & \frac{\cos(\theta)\cos(\varphi)}{\cos(\theta-\varphi)}\;, \\
    \label{eq:coefficien_of_H1}
    |\alpha| & = & \sqrt{-\frac{\tan(\theta)\tan(\varphi)\csc^{2}(\theta-\varphi)}{(\tan(\theta)\tan(\varphi)+1)^{2}}}=\frac{\sqrt{-\sin(2\theta)\sin(2\varphi)}}{|\sin(2(\theta-\varphi))|} \;.
\end{eqnarray}

\subsection*{Complex $H_1$}

A complex $H_1$ can be parameterized as
\begin{equation}
    \begin{aligned}
        H_1 & =\alpha\vert\theta,\delta\rangle\langle\varphi,\gamma|=\alpha\left(\begin{array}{cc}
            \cos\theta\cos\varphi & e^{i\gamma}\cos\theta\sin\varphi\\
            e^{-i\delta}\sin\theta\cos\varphi & e^{-i\left(\delta-\gamma\right)}\sin\theta\sin\varphi
        \end{array}\right) \;, \\
        \vert\varphi,\gamma\rangle & =\left(\begin{array}{c}
            \cos\varphi\\
            e^{i\gamma}\sin\varphi
        \end{array}\right) \;, \\
        \vert\theta,\delta\rangle & =\left(\begin{array}{c}
            \cos\theta\\
            e^{i\delta}\sin\theta
        \end{array}\right) \;,
    \end{aligned}
	\label{eq:complex H1}
\end{equation}
where $\alpha$ is a complex number.

Following the same procedure as for real $H_1$ we obtain
\begin{align}
    \begin{aligned}
        \efb & =\frac{e^{i\delta}\cot(\theta)\cos(\varphi)}{e^{i\gamma}\sin(\varphi)+e^{i\delta}\cot(\theta)\cos(\varphi)} \;, \\
        |\alpha| & =\frac{2e^{2i(\gamma+\delta)}\sin(2\theta)\sin(2\varphi)}{\left(\left(e^{i\gamma}-e^{i\delta}\right)^{2} (-\cos(2(\theta+\varphi)))+\left(e^{i\gamma}+e^{i\delta}\right)^{2}\cos(2(\theta-\varphi))-4e^{i(\gamma+\delta)}\right)\left(e^{i\gamma}\cos(\theta)\cos(\varphi)+e^{i\delta}\sin(\theta)\sin(\varphi)\right)^{2}} \;.
    \end{aligned}
\label{sol_for_comp}
\end{align}
Reality of $|\alpha|$ imposes $\delta=\gamma$, and consequently
\begin{eqnarray}
    \label{eq:eigenvalue_complex}
    \efb & = & \frac{\cos(\theta)\cos(\varphi)}{\cos(\theta-\varphi)}\;, \\
    \label{eq:coefficient_complex}
    |\alpha| & = & \frac{\sqrt{- \sin(2\theta)\sin(2\varphi)}}{|\sin(2(\theta-\varphi))|} \;.
\end{eqnarray}

Solutions~\eqref{eq:eigen_value},~\eqref{eq:coefficien_of_H1} and~\eqref{eq:eigenvalue_complex},~\eqref{eq:coefficient_complex} are identical. Since $|\alpha|$ is real, solutions exist only in the parameter regions $0\le\theta\le\frac{\pi}{2}\cap\frac{\pi}{2}\le\varphi\le\pi$ or $\frac{\pi}{2}\le\theta\le\pi\cap0\le\varphi\le\frac{\pi}{2}$.

In Bloch representation the Hamiltonian reads
\begin{equation}
    H_{k}=H_{1}^{\dagger}e^{ik}+H_{0}+H_{1}e^{-ik} \;.
    \label{eq:bloch eq}
\end{equation}
With the above parameterization~\eqref{eq:complex H1} and~\eqref{eq:canonical-H0} the band structure follows as
\begin{equation}
    \begin{aligned}
        E_{FB} & =\frac{\cos\theta\cos\varphi}{\cos(\theta-\varphi)} \;, \\
        E_k & =\frac{\cos\theta\cos\varphi}{\cos(\theta-\varphi)}+2|\alpha|\cos(\theta-\varphi)\cos(k+\phi_{\alpha})\;,
    \end{aligned}
    \label{eq:band-struct}
\end{equation}
where $\phi_{\alpha}$ is the phase of $\alpha=|\alpha|e^{i\phi_{\alpha}}$.

\subsection*{Degenerate $H_0$ }

The solutions $\alpha$ and $\efb$ in~\eqref{eq:eigenvalue_complex} and~\eqref{eq:coefficient_complex} diverge for $\theta-\varphi=\pm\frac{\pi}{2}$ and $\theta=\phi$, and so does $H_1$. We renormalize the Hamiltonian by multiplying it with $\frac{1}{\alpha}$. Then $H_0$ vanishes, and $H_1$ turns finite. 

However, when $\theta-\varphi=\pm\frac{\pi}{2}$ the dispersion bandwidth in~\eqref{eq:band-struct}) is finite, and after normalization the dispersive band becomes flat as well. Therefore we have two coexisting flatbands on the lines $\theta-\varphi=\pm\frac{\pi}{2}$. According to~\eqref{eq:redef_cls_con1} and~\eqref{eq:redef_cls_con2} $\vec{\psi}_1\perp\vec{\psi}_2$. In such a case we can always perform a rotation in each unit cell such that $\psi_1=(1,0)$, $\psi_2=(0,1)$. A subsequent redefinition of the unit cell turns the CLS into a $U=1$ class one, see above subsection
on orthogonality for $\nu=2$ and $U=2$.

When $\theta=\varphi$, according to~\eqref{eq:redef_cls_con1} and~\eqref{eq:redef_cls_con2}) $\vec{\psi}_1\parallel\vec{\psi}_2$, therefore linearly dependent, and the flat band turns into the $U=1$ class. In this case 
\begin{equation}
    H_1=\vert\theta,\gamma\rangle\langle\theta,\gamma\vert=
    \left(\begin{array}{cc}
        2\cos^{2}\theta & \frac{1}{2}e^{i\gamma}\sin(2\theta)\\
        \frac{1}{2}e^{-i\gamma}\sin(2\theta) & 2\cos^{2}\theta
    \end{array}\right) \;.
\end{equation}
The corresponding Bloch Hamiltonian in momentum space reads
\begin{equation}
    H_k = H_1^\dagger e^{ik} + H_1 e^{-ik}=
    \cos k\left(\begin{array}{cc}
        2\cos^{2}\theta & e^{i\gamma}\sin(2\theta)\\
        e^{-i\gamma}\sin(2\theta) & 2\cos^{2}\theta
    \end{array}\right) \;,
\end{equation}
which yields one flat and one dispersive band 
\begin{equation}
    E_{FB} = 0 \;,\;  E_k=2\cos k \;.
\end{equation}

\subsection*{FB energy equals one of the eigenvalues of $H_0$: reduction to $U=1$}

When the flatband energy $\efb$ equals to one of the eigenvalues $0,\ 1$ of $H_0$, we have to solve the original equations~(\ref{eq:eigenvalue_eq1}-\ref{eq:u2-cls-con2}). A simple calculation shows that the only remaining flatband solutions are again of class $U=1$.

\section{Generalized sawtooth chain}

In our FB generator we have considered the canonical form of $H_0$, which was diagonal. We can perform unitary transformations (rotations) of the unit cell basis
which will modify $H_1$ and make $H_0$ non-diagonal, turning the whole model into a non-canonical one.
Using the rotation matrix
\begin{equation}
	R\left(\omega\right) = 
	\left(\begin{array}{cc}
		\cos\omega & -\sin\omega\\
		\sin\omega & \cos\omega
	\end{array}\right),
\end{equation}
we define $\tilde{\psi}_i=R(\omega) \vec{\psi}_i$, $\tilde{H}_m=R(\omega)H_m R^\dagger(\omega)$, where $m=-1,0,1$ and $H_{-m}=H_m^\dagger$ and
\begin{equation}
	H_0 = \left(\begin{array}{cc}
		0 & 0\\
		0 & 1
	\end{array}\right),\ \ 
	H_1 = \alpha\left(\begin{array}{cc}
		\cos\theta\cos\varphi & e^{i\gamma}\cos\theta\sin\varphi\\
		e^{-i\gamma}\sin\theta\cos\varphi & \sin\theta\sin\varphi
	\end{array}\right)
\end{equation}
with $|\alpha|=\frac{\sqrt{-\sin(2\theta)\sin(2\varphi)}}{|\sin(2(\theta-\varphi))|}$. We consider the case $\gamma=\delta=0$, thus $H_0$ and $H_1$ read
\begin{equation}
	\begin{aligned}
		\tilde{H_0} & = R^\dagger(\omega) H_0 R(\omega) = \left(\begin{array}{cc}
			\sin^2\omega & \cos\omega\sin\omega\\
			\cos\omega\sin\omega & \cos^{2}\omega
		\end{array}\right)\;,\\
		\tilde{H_1} & = \alpha\left(\begin{array}{cc}
			\cos(\theta+\omega)\cos(\varphi+\omega) & \cos(\theta+\omega)\sin(\varphi+\omega)\\
			\cos(\varphi+\omega)\sin(\theta+\omega) & \sin(\theta+\omega)\sin(\varphi+\omega)
		\end{array}\right)\;.
	\end{aligned}
\label{34}
\end{equation}
We can always find a value of $\omega$ which will zero one row or one column of $H_1$. This simplifies the non-canonical lattice into a {\sl generalized sawtooth} chain.
It has in general  three different hopping strengths and an onsite energy detuning between the two sites in a unit cell.
As an example we consider the case when the first column of $H_1$ vanishes:
\begin{equation}
	\varphi + \omega = \pm\frac{\pi}{2}\;.
\end{equation}
It follows
\begin{equation}
	\begin{aligned}
		\tilde{H}_0 &= \left(\begin{array}{cc}
			\cos ^2(\varphi ) & -\cos (\varphi ) \sin (\varphi ) \\
			-\cos (\varphi ) \sin (\varphi ) & \sin ^2(\varphi ) \\
		\end{array}\right) = \left(\begin{array}{cc}
			\epsilon_1 & t_1\\
			t_1 & \epsilon_2
		\end{array}\right)  \;, \\
		\tilde{H}_1 &= \alpha\left(\begin{array}{cc}
			0 & -\sin(\theta-\varphi)\\
			0 & \cos(\theta-\varphi)
		\end{array}\right) = \left(\begin{array}{cc}
			0 & t_2\\
			0 & t_3
		\end{array}\right) \;.
	\end{aligned}
\end{equation}
For the particular case of the ST1 chain (sawtooth chain with two hoppings equal, and zero onsite energy detuning, see main text)
$\epsilon_1=\epsilon_2$ and $t_1=t_2$, we find
\begin{equation}
	\begin{aligned}
		\varphi &= \frac{3\pi}{4}, \ \ \theta = \arctan(3+2\sqrt{2}) \;,  \\
		\varphi &= \frac{\pi}{4}, \ \ \theta = \pi - \arctan(3-2\sqrt{2}).
	\end{aligned}
\end{equation}
This leads to the following tight-binding equations:
\begin{equation}
Ea_n = -\sqrt{2} b_n - \sqrt{2} b_{n+1} \;,\; E b_n =-\sqrt{2} a_n -\sqrt{2} a_{n-1} -b_{n+1} - b_{n-1} \;.
\end{equation}
The flat band is located at $E_{FB}=2$ (Fig.1(b) in the main text). A compact localized eigenstate has the form $a_0=a_1=1$ and $b_1=-\sqrt{2}$ (up to a normalization factor) with all other amplitudes vanishing.

Detuning the angles $\theta,\varphi$ away from this point we deform the ST1 model, while maintaining one band flat.

Let us require $t_1=t_2=t_3$. It follows
\begin{equation}
	\begin{aligned}
		\theta &= \frac{\pi }{2} - \frac{1}{2} \tan^{-1}\left(\frac{1}{2}\right), \ \ \varphi = \frac{3 \pi }{4} - \frac{1}{2} \tan^{-1}\left(\frac{1}{2}\right) \;, \\
		\theta &= \frac{\pi }{2} - \frac{1}{2} \tan^{-1}\left(\frac{1}{2}\right), \ \ \varphi = \frac{3 \pi }{4} - \frac{1}{2} \tan^{-1}\left(\frac{1}{2}\right) \;.
	\end{aligned}
\end{equation}
This is a novel high symmetry sawtooth chain (ST2 chain, see main text). It can be obtained e.g. with the following matrices 
\begin{equation}
	\begin{aligned}
		H_0 = - \left(\begin{array}{cc}
			0 & 1\\
			1 & 1
		\end{array}\right),\  &
		H_1 = -\left(\begin{array}{cc}
			0 & 1\\
			0 & 1
		\end{array}\right).
	\end{aligned}
\end{equation}
This leads to the following tight-binding equations:
\begin{equation}
Ea_n = -b_n - b_{n+1} \;,\; E b_n = -b_n -a_n -a_{n-1} -b_{n+1} - b_{n-1} \;.
\end{equation}
The flat band is located at $E_{FB}=1$ (Fig.1(c) in the main text). A compact localized eigenstate has the simple form $a_0=a_1=-b_1=1$ (up to a normalization factor) with all other amplitudes vanishing.
Note that we can also set other columns and rows of $H_1$ in (\ref{34}) to zero, and obtain all points in the $\theta, \varphi$ diagram corresponding to ST1 and ST2 chains. All these points are shown by filled squares and filled circles in Fig. 3 in the main text.

\end{document}